\documentclass[12pt]{article}

\usepackage[utf8]{inputenc}
\usepackage[english]{babel}
\usepackage{epsfig}
\usepackage{amsmath}
\usepackage{amssymb}
\usepackage{color}
\usepackage{bbm}
\usepackage{hyperref}
\usepackage{physics}
\usepackage{graphicx}
\usepackage{xcolor,colortbl}
\usepackage{underscore}
\usepackage{cite}
\usepackage{listings}

\newcommand{\be}{\begin{equation}}
\newcommand{\ee}{\end{equation}}
\newcommand{\ba}{\begin{eqnarray}}
\newcommand{\ea}{\end{eqnarray}}
\newcommand{\bs}{\begin{subequations}}
\newcommand{\es}{\end{subequations}}
\newcommand{\no}{\nonumber\\}

\def\Math{{\tt \textsc{Mathematica} }}
\def\SW{{\tt \textsc{StableWein}}}

\makeatletter
\def\hlinewd#1{%
	\noalign{\ifnum0=`}\fi\hrule \@height #1 %
	\futurelet\reserved@a\@xhline}
\makeatother
 
\newcommand{\blue}{\color{blue!90!black}}
\newcommand{\red}{\color{red!90!black}}
\newcommand{\green}{\color{green!60!black}}

\lstdefinelanguage{Math14}
{morekeywords={},
sensitive=false,
morecomment=[l]{//},
morecomment=[s]{(*}{*)},
morestring=[b]",
}

\definecolor{codegreen}{rgb}{0,0.6,0}
\definecolor{codegray}{rgb}{0.95,0.95,0.95}
\definecolor{codepurple}{rgb}{0.0,0,0.8}
\definecolor{backcolour}{rgb}{0.95,0.75,0.92}
\definecolor{codecyan}{rgb}{0.29, 0.62, 0.73}
\definecolor{codemagenta}{rgb}{0.95,0.0,0.0}

\lstdefinestyle{mystyle}{
    backgroundcolor=\color{codegray},   
    commentstyle=\color{codecyan},
    keywordstyle=\color{codepurple},
    stringstyle=\color{gray!70!black},    
    basicstyle=\ttfamily\footnotesize,
    breakatwhitespace=false,         
    breaklines=true,                 
    captionpos=b,                    
    keepspaces=false,                 
    showspaces=false,                
    showstringspaces=false,
    showtabs=false,                  
    tabsize=2,
    lineskip=0.8ex
}

\graphicspath{ {figures/} }

\begin{document}

\title{\LARGE
  Assessing boundedness from below
  in the $\mathbb{Z}_2 \times \mathbb{Z}_2$-symmetric
  three-Higgs-doublet model: algorithm and machine learning}

\author{
  Darius~Jur\v{c}iukonis,$^{(1)}$\thanks{E-mail:
    \tt darius.jurciukonis@tfai.vu.lt}
  \
  Lu\'\i s~Lavoura,$^{(2)}$\thanks{E-mail:
    \tt balio@cftp.tecnico.ulisboa.pt}
  \ and
  Andr\'e Milagre$^{(2)}$\thanks{E-mail:
    \tt andre.milagre@tecnico.ulisboa.pt}
  \
  \\*[3mm]
  $^{(1)}\!$
  \small Vilnius University, Institute of Theoretical Physics and Astronomy, \\
  \small Saul\.etekio~av.~3, Vilnius 10257, Lithuania
  \\*[2mm]
  $^{(2)}\!$
  \small Universidade de Lisboa, Instituto Superior T\'ecnico, CFTP, \\
  \small Av.~Rovisco~Pais~1, 1049-001~Lisboa, Portugal
}

\maketitle

\begin{abstract}
  The scalar potential of any particle-physics model
  must be bounded from below (BFB).
  We consider the extension of the Standard electroweak Model
  with three $SU(2)$ doublets of scalars
  and a symmetry under which each of those doublets changes sign.
  In the absence of necessary and sufficient conditions
  for boundedness from below (BnessFB) for this specific model,
  we argue that one may use
  \textcolor{black}{increasingly stringent sets of necessary conditions.}
  We introduce a \Math code,
  \SW,
  that implements this idea.
  The user is allowed to choose the level of accuracy 
  that they want in the determination of BnessFB;
  more precision means the use of more necessary conditions,
  and usually entails a longer running time for the code.
  Our investigation suggests that our procedure and code can be
  extremely precise in the determination of the potentials that are BFB.
  In addition,
  we introduce a machine-learning code that identifies,
  with more than 99\% accuracy,
  which potentials are BFB.
\end{abstract}

\section{Introduction}

Throughout this paper,
the renormalizable scalar potentials are of the form $V_2 + V_4$,
where $V_2$ is quadratic in the scalar fields
and $V_4$ is quartic in the scalar fields.

In the extremely successful Standard Model (SM)
of the electroweak interactions,
the gauge group is $SU(2) \times U(1)$
and there is only one $SU(2)$ doublet of scalars:
$\Phi$.
Hence,
$V_2 = \mu\, \Phi^\dagger \Phi$
and $V_4 = \lambda \left( \Phi^\dagger \Phi \right)^2$,
where $\mu$ and $\lambda$ are couplings,
\textit{i.e.}\ numbers (not fields).

The consistency of any particle-physics model
necessitates that its scalar potential is bounded from below (BFB).
This means that the potential
can never tend to minus infinity---else the Hamiltonian does not have a minimum,
\textit{i.e.}\ the theory has no vacuum state.
Boundedness from below (BnessFB) is equivalent to $V_4$
\textcolor{black}{being positive}
for any values of the scalar fields.\footnote{\textcolor{black}{When one uses
  our condition that $V_4$ be \emph{positive}
  for any values of the scalar fields,
  $V_2$ is irrelevant for BnessFB.
  The weaker condition where $V_4$ is allowed to be zero
  necessitates that one furthermore makes sure that
  $V_2 \ge 0$ whenever $V_4 = 0$.
  This may be tricky~\cite{tooby},
  therefore we prefer the stricter condition $V_4 > 0$.}
}
\textcolor{black}{If $V_4$ were negative
  for some values of the scalar fields, then}
multiplying those values by an ever-increasing positive number
would make the potential tend to minus infinity.
In the SM,
since $\Phi^\dagger \Phi \ge 0$,
the potential is BFB if and only if $\lambda$ is 
\textcolor{black}{positive}.

For a variety of reasons---like accommodating nonzero neutrino masses,
dark matter,
and baryogenesis\footnote{In the SM there is neither sufficient $CP$ violation
nor the first-order phase transition necessary
for baryogenesis.}---many particle physicists
like to extend the SM.
Most extensions have a scalar sector larger than the one of the SM,
in particular more than one $SU(2)$ doublet.
(There are also extensions of the SM with a larger gauge group.)
The scalar sector of the two-Higgs-doublet model (2HDM)~\cite{sher}
has been extensively explored.
This paper concentrates on the model with three scalar $SU(2)$ doublets $\Phi_k$
($k = 1, 2, 3$ throughout this paper),
which is called the three-Higgs-doublet model (3HDM).\footnote{There are many
3HDMs,
depending both on the additional symmetries that are imposed and
on the fermionic sector.}

Unfortunately,
for all but the simplest extensions of the SM
$V_4$ has many couplings---not just one,
like $\lambda$ in the SM---and it is not possible
to establish mathematical necessary and sufficient conditions for BnessFB
in terms of those couplings alone.
This has been achieved~\cite{silva} for the 2HDM,
where $V_4$ has $2^2 \left( 2^2 + 1 \right) / 2 = 10$ real couplings;
for the 3HDM,
where $V_4$ has $3^2 \left( 3^2 + 1 \right) / 2 = 45$
real couplings~\cite{BentoHilbert},
it could not be achieved.

However,
most 3HDMs used in practical models have additional (non-gauge,
or `internal') symmetries.
Those symmetries restrict $V_4$,
making it have much less than 45 real couplings.
The symmetries that the scalar potential of a 3HDM may enjoy
have been classified. 
Some of them have a $U(1)_1 \times U(1)_2$ subgroup,
where
\bs
\label{u1u1}
\ba
U(1)_1: & & \Phi_2 \to e^{i \varrho} \Phi_2,\ \Phi_3 \to e^{- i \varrho} \Phi_3;
\label{u11}
\\*[1mm]
U(1)_2: & & \Phi_1 \to e^{2 i \varsigma} \Phi_1,\
\Phi_2 \to e^{- i \varsigma} \Phi_2,\
\Phi_3 \to e^{- i \varsigma} \Phi_3.
\label{u12}
\ea
\es
In Eqs.~\eqref{u1u1},
$\varrho$ and $\varsigma$ are arbitrary phases. 
Other internal symmetries
have just a $\mathbb{Z}_2^{(3)} \times U(1)_2$ subgroup,
where $\mathbb{Z}_2^{(3)}$ is generated by the transformation
\be
\label{z23}
\mathbb{Z}_2^{(3)}: \quad
\Phi_3 \to - \Phi_3.
\ee
Still other internal symmetries have only
a $\mathbb{Z}_2^{(3)} \times \mathbb{Z}_2^{(2)}$
subgroup,\footnote{The 3HDM firstly (to our knowledge) used in particle physics
had $\mathbb{Z}_2^{(3)} \times \mathbb{Z}_2^{(2)}$
symmetry~\cite{weinberg,deshpande}.
Inspired by Ref.~\cite{weinberg},
we call `Weinberg model' to a 3HDM
with $\mathbb{Z}_2^{(3)} \times \mathbb{Z}_2^{(2)}$ symmetry.
This,
together with some authors' use of designating the conditions for BnessFB
as `vacuum stability conditions',
is the origin of the name \SW\
of the package introduced in this paper.
The $\mathbb{Z}_2^{(3)} \times \mathbb{Z}_2^{(2)}$-symmetric 3HDM
furnished with (spontaneously broken) $CP$ symmetry
was pioneered in Ref.~\cite{branco};
we call it 
`Branco model'.}
where $\mathbb{Z}_2^{(2)}$ is generated by the transformation
\be
\label{z22}
\mathbb{Z}_2^{(2)}: \quad
\Phi_2 \to - \Phi_2.
\ee
Finally,
there are internal symmetries that do not have any of these subgroups.
All possible internal symmetries of the renormalizable potential of a 3HDM
are listed in Appendix~\ref{app:symmetries}.

So,
usually one does not need to establish whether the potential of the full 3HDM
is BFB---one only has to do it for a $V_4$ constrained by a symmetry,
hence with less than 45 real couplings.
Notably,
a mathematical algorithm for deciding on the BnessFB
of a potential with $U(1)_1 \times U(1)_2$ symmetry
has been devised~\cite{faro1}
and extended to a potential
with $\mathbb{Z}_2^{(3)} \times U(1)_2$ symmetry~\cite{faro2}.
The problem of the BnessFB of a potential
with only $\mathbb{Z}_2^{(3)} \times \mathbb{Z}_2^{(2)}$ symmetry
remains extant.\footnote{The exception are the potentials
with either $S_4$ or $SO(3)$ symmetries~\cite{vazao,buskin}
(see the third column of Table~\ref{table:symmetries}).
\textcolor{black}{However,
  those symmetries are much stronger
  than just $\mathbb{Z}_2^{(3)} \times \mathbb{Z}_2^{(2)}$.
}
}

Sufficient conditions for BnessFB
of a $\mathbb{Z}_2^{(3)} \times \mathbb{Z}_2^{(2)}$-symmetric 3HDM potential
have been devised~\cite{GOO,boto}.
They have the obvious setback that one will be exploring only
the portion of the space of BFB potentials
that fulfill the sufficient conditions,
neglecting other,
potentially physically interesting parts of that space.
In this paper, we contend that,
in the case of $\mathbb{Z}_2^{(3)} \times \mathbb{Z}_2^{(2)}$-symmetric
potentials,
it is possible to take the alternative approach of using \emph{necessary},
instead of sufficient,
BnessFB conditions.
In our approach, one incurs the risk of using potentials
that fulfill the necessary conditions but are not really BFB;
however,
that risk may be systematically curtailed by using more and more
necessary conditions,
as we shall illustrate in this paper,
thus diminishing the number of non-BFB potentials that one admits for analysis.
Our explorations suggest that it is possible,
without excessive computational effort,
to utilize a set of necessary conditions that decides,
\emph{with extremely high accuracy} (much above 99\%),
which $\mathbb{Z}_2^{(3)} \times \mathbb{Z}_2^{(2)}$-symmetric 3HDM potentials
are BFB.

This paper introduces the \Math package \SW\
that encodes the result of our findings.
The user of that notebook inputs the quartic part
of a $\mathbb{Z}_2^{(3)} \times \mathbb{Z}_2^{(2)}$-symmetric potential
and receives as output the indication of whether that potential is
(likely to be)
BFB or not.
The level of precision of that indication may be chosen by the user:
a greater level of precision means that
more necessary conditions are employed and,
consequently,
the code takes a longer time to reach the indication.
One further option of our code minimizes $V_4$ through brute force,
in principle obtaining---if the minimization has been well done---the
exact answer to the BnessFB question for that potential.
Our code may be employed either for $CP$-violating
$\mathbb{Z}_2^{(3)} \times \mathbb{Z}_2^{(2)}$-symmetric potentials
or for $CP$-conserving ones;
in the two cases the code uses somewhat different necessary conditions,
and it achieves greater precision in the $CP$-conserving case
(which has one parameter less in $V_4$).

Additionally,
this paper also introduces neural networks (NNs)\footnote{See
Ref.~\cite{khanna} for a paper where an NN is used
to analyze a 3HDM, albeit one with $\mathbb{Z}_3$ instead of
$\mathbb{Z}_2 \times \mathbb{Z}_2$ symmetry and with emphasis
elsewhere rather than on the BFB constraints on $V_4$.}
that have
been trained to identify BFB
$\mathbb{Z}_2^{(3)} \times \mathbb{Z}_2^{(2)}$-symmetric potentials;
they too are implemented in the \SW\ package.
This artificial intelligence,
which uses neither sufficient nor necessary BFB conditions,
displays an accuracy close to 99.9\%
in deciding which potentials are BFB.

This paper is organized as follows.
In Sec.~\ref{sec:necCond} we derive necessary conditions
for BnessFB of a $\mathbb{Z}_2 \times \mathbb{Z}_2$-invariant 3HDM potential.
In Sec.~\ref{sec:Branco's} we particularize the conditions
of Sec.~\ref{sec:necCond} to a $CP$-conserving potential.
In Sec.~\ref{sec:sufficient} we provide sufficient conditions for BnessFB.
In Sec.~\ref{sec:minimization} we describe the minimization procedures used.
In Sec.~\ref{sec:package} we give instructions for using the package \SW.
The main conclusions of this paper
are summarized in Sec.~\ref{sec:conclusions}. 
Two appendices,
which may be altogether omitted,
deal on known material:
Appendix~\ref{app:symmetries} lists the possible symmetries of a 3HDM;
Appendix~\ref{app:unitarity} reviews the
perturbative unitarity constraints
for the $\mathbb{Z}_2 \times \mathbb{Z}_2$-invariant $V_4$.

\section{The necessary conditions for BnessFB}
\label{sec:necCond}

We refer to the $\mathbb{Z}_2 \times \mathbb{Z}_2$-symmetric 3HDM
as Weinberg model.
In this section,
we define the three $SU(2)$ scalar doublets
together with the quartic part of the scalar potential $V_4$.
We introduce the phase space coordinates $\varpi_k$ and $\vartheta_k$
and present the copositivity condition on $V_4$,
from which we derive several sets of necessary conditions
with increasing levels of accuracy.
Throughout, we use $k=1,2,3$.

\paragraph{Doublets:} We write
\be
\label{jbp}
\Phi_1 = \left( \begin{array}{c} a \\ b \end{array} \right), \qquad
\Phi_2 = \left( \begin{array}{c} c \\ d \end{array} \right), \qquad
\Phi_3 = \left( \begin{array}{c} e \\ f \end{array} \right),
\ee
where $a, \ldots, f$ are complex scalar (Klein--Gordon) fields.
Under an $SU(2) \times U(1)$ transformation
$\Phi_k \to U_2 \Phi_k\ \forall k$,
where $U_2$ is a matrix of $U(2)$.\footnote{Note that
$U(2) \cong \left. \left[ SU(2) \times U(1) \right] \right/ \mathbb{Z}_2$
is locally identical to $SU(2) \times U(1)$,
but the two groups have different topologies.
See for instance Ref.~\cite{vicente}.}
Without loss of generality, it is possible to make $a = 0$
everywhere in space--time through a (local) $SU(2)$ transformation,
and then \textcolor{black}{to make both $b$ and $c$ real
  through two (local) $U(1)$ transformations---one of them
  being the $U(1)$ in $SU(2) \times U(1)$ and the other one being
  the neutral component of $SU(2)$.}

\paragraph{Parameters $\varpi_k$ and $\vartheta_k$:}
Assuming $\Phi_k^\dagger \Phi_k \ne 0\ \forall k$,\footnote{If
$\exists k: \Phi_k^\dagger \Phi_k = 0$,
then the 3HDM reduces to a 2HDM,
and the BFB conditions for a 2HDM are known.
See below.}
we define
\be
\label{pis}
\pi_1 := \frac{\left( \Phi_2^\dagger \Phi_3 \right)^2}{\Phi_2^\dagger \Phi_2\,
  \Phi_3^\dagger \Phi_3}, \qquad
\pi_2 := \frac{\left( \Phi_3^\dagger \Phi_1 \right)^2}{\Phi_3^\dagger \Phi_3\,
  \Phi_1^\dagger \Phi_1}, \qquad
\pi_3 := \frac{\left( \Phi_1^\dagger \Phi_2 \right)^2}{\Phi_1^\dagger \Phi_1\,
  \Phi_2^\dagger \Phi_2}.
\ee
We moreover define
\be
\label{varpi}
\varpi_k := \left| \pi_k \right|, \qquad
\vartheta_k := \arg \pi_k, \qquad
\vartheta := \sum_{k=1}^3 \vartheta_k.
\ee
The
$\varpi_k$ satisfy
\be
0 \le \varpi_k \le 1.
\ee
The six dimensionless parameters $\varpi_k$ and $\vartheta_k$
are not independent,
because
\be
\label{mbp}
2 \left( 1 + \cos \vartheta \right) \prod_{k=1}^3 \varpi_k
= \left( 1 - \sum_{k=1}^3 \varpi_k \right)^2.
\ee
Equation~\eqref{mbp} has been derived in Ref.~\cite{vazao}
and follows from substituting Eqs.~\eqref{jbp}
in Eqs.~\eqref{pis}.\footnote{Equation~\eqref{mbp}
would not apply if the $\Phi_k$ were triplets of $SU(3)$
instead of doublets of $SU(2)$.
On the other hand,
if the $\Phi_k$ were singlets instead of doublets,
then the much stronger equations $\vartheta = 0$
and $\varpi_k = 1\ \forall k$ would hold.}
It implies that $\cos{\vartheta} \ge 0$
and leaves the sign of $\vartheta$ un-determined.
Because of Eq.~\eqref{mbp},
\be
\label{cbe}
4\, \prod_{k=1}^3 \varpi_k \ge \left( 1 - \sum_{k=1}^3 \varpi_k \right)^2,
\ee
which constrains only the $\varpi_k$.

\begin{figure}[h!] 
 \centering
  \begin{tabular}{c}
	\epsfig{file=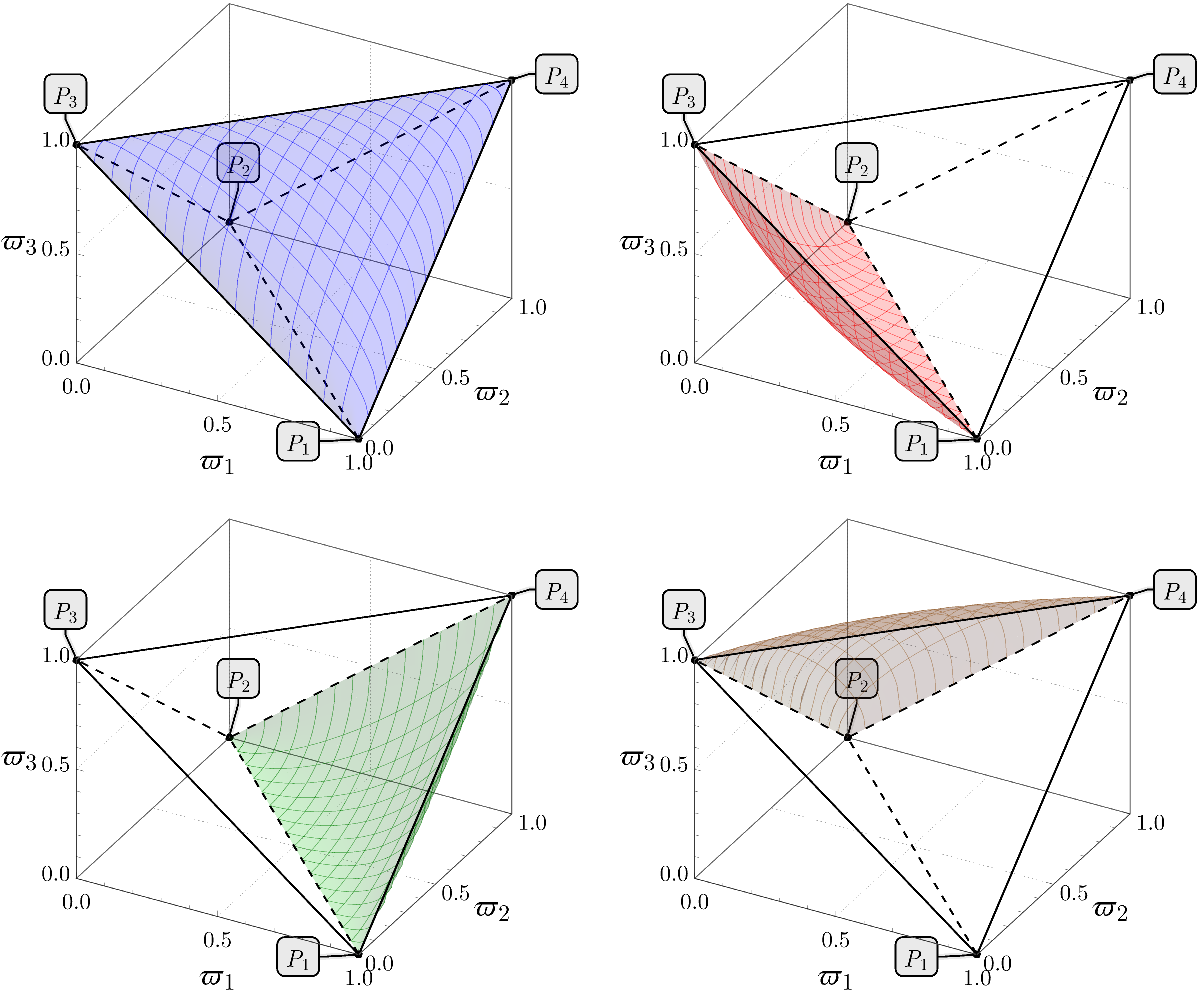,width=0.9\textwidth}
  \end{tabular}
  \caption{A perspective of the allowed volume
    in $\left( \varpi_1, \varpi_2, \varpi_3 \right)$ space.
    The four points $P_{1,2,3,4}$ defined in Eqs.~\eqref{points} are displayed.
    The three edges connecting $P_1$,
    $P_3$,
    and $P_4$ are displayed as full straight lines;
    the edges connecting each of those three points to $P_2$ are dashed lines.
    In each of the four panels of the figure,
    the four faces of the surface that bounds the allowed volume
    are separately displayed in different colours;
    in each case,
    a darker tone colours the part of the face that is visible
    in this perspective,
    and a lighter tone marks the part of the face that is hidden
    behind the volume.
    The four faces are identical and convex.}
  \label{fig:allowed}
\end{figure}

\paragraph{Special points:} Using the notation
$\left( \varpi_1, \varpi_2, \varpi_3 \right)$,
the four points
\be
\label{points}
P_1 = \left( 1, 0, 0 \right), \qquad P_2 = \left( 0, 1, 0 \right), \qquad
P_3 = \left( 0, 0, 1 \right), \qquad P_4 = \left( 1, 1, 1 \right)
\ee
saturate the inequality~\eqref{cbe}.
They are the corners of the allowed volume
in $\left( \varpi_1, \varpi_2, \varpi_3 \right)$ space,
which is displayed in Fig.~\ref{fig:allowed}.
Point $P_1$ is reached,
for instance,
when $a = d = f = 0$;
point $P_4$ is reached when $a = c = e = 0$
(\textit{i.e.}\ when all three doublets
only have `neutral' components~\cite{faro1} or are equivalent to singlets).
Using Eq.~\eqref{mbp} one sees that $\vartheta = 0$ at point $P_4$,
while points $P_k$ have un-determined $\vartheta$.
As a matter of fact,
$\vartheta$ is un-determined at the three edges connecting $P_1$,
$P_2$,
and $P_3$,
and $\vartheta$ is zero
at the whole rest of the surface that bounds
the allowed $\left( \varpi_1, \varpi_2, \varpi_3 \right)$ volume.
At the center of the allowed volume,
given by $\varpi_k = 1/2\ \forall k$,
$\cos{\vartheta} = 0$.

\paragraph{Potential:}
The $\mathbb{Z}_2^{(3)} \times \mathbb{Z}_2^{(2)}$-symmetric 3HDM
has quartic potential
\bs
\label{uty}
\ba
V_4 &=& \sum_{k=1}^3 \frac{a_k}{2} \left( \Phi_k^\dagger \Phi_k \right)^2
\\ & &
+ b_1\, \Phi_2^\dagger \Phi_2\, \Phi_3^\dagger \Phi_3
+ b_2\, \Phi_3^\dagger \Phi_3\, \Phi_1^\dagger \Phi_1
+ b_3\, \Phi_1^\dagger \Phi_1\, \Phi_2^\dagger \Phi_2
\\ & &
+ c_1\, \Phi_2^\dagger \Phi_3\, \Phi_3^\dagger \Phi_2
+ c_2\, \Phi_3^\dagger \Phi_1\, \Phi_1^\dagger \Phi_3
+ c_3\, \Phi_1^\dagger \Phi_2\, \Phi_2^\dagger \Phi_1
\\ & &
+ \left[
  e^{i \epsilon_1}\, \frac{d_1}{2} \left( \Phi_2^\dagger \Phi_3 \right)^2
  + e^{i \epsilon_2}\, \frac{d_2}{2} \left( \Phi_3^\dagger \Phi_1 \right)^2
  + e^{i \epsilon_3}\, \frac{d_3}{2} \left( \Phi_1^\dagger \Phi_2 \right)^2
  + \mathrm{H.c.} \right],
\label{jbu}
\ea
\es
with twelve real couplings $a_k$,
$b_k$,
$c_k$,
and $d_k$.
The three phases $\epsilon_k$ are not all physical;
only
\be
\epsilon := \sum_{k=1}^3 \epsilon_k
\ee
has physical significance.
Indeed,
by changing the relative phases of the three doublets
via $U(1)$ transformations,
one may change the three $\epsilon_k$,
but $\epsilon$ remains invariant.
In particular,
one may re-write line~\eqref{jbu} in any one of the following three forms:
\bs
\label{three_forms}
\ba
& & + \left[
  e^{i \epsilon}\, \frac{d_1}{2} \left( \Phi_2^\dagger \Phi_3 \right)^2
  + \frac{d_2}{2} \left( \Phi_3^\dagger \Phi_1 \right)^2
  + \frac{d_3}{2} \left( \Phi_1^\dagger \Phi_2 \right)^2
  + \mathrm{H.c.} \right];
\label{hjv} \\
& & + \left[
  \frac{d_1}{2} \left( \Phi_2^\dagger \Phi_3 \right)^2
  + e^{i \epsilon}\, \frac{d_2}{2} \left( \Phi_3^\dagger \Phi_1 \right)^2
  + \frac{d_3}{2} \left( \Phi_1^\dagger \Phi_2 \right)^2
  + \mathrm{H.c.} \right];
\\
& & + \left[
  \frac{d_1}{2} \left( \Phi_2^\dagger \Phi_3 \right)^2
  + \frac{d_2}{2} \left( \Phi_3^\dagger \Phi_1 \right)^2
  + e^{i \epsilon}\, \frac{d_3}{2} \left( \Phi_1^\dagger \Phi_2 \right)^2
  + \mathrm{H.c.} \right].
\ea
\es
Thus,
$V_4$ has 13 real couplings---the $a_k$,
$b_k$,
$c_k$,
$d_k$,
and $\epsilon$---as advertised in the previous section.
The perturbative unitarity bounds on those 13 parameters are given
in Appendix~\ref{app:unitarity};
they have been implemented in \SW.

\paragraph{NCL1:}
Using the definitions~\eqref{pis} and~\eqref{varpi},
one may re-write Eq.~\eqref{uty} as
\be
\label{mvo}
V_4 = \frac{1}{2} \left( \begin{array}{ccc}
  \Phi_1^\dagger \Phi_1, &
  \Phi_2^\dagger \Phi_2, &
  \Phi_3^\dagger \Phi_3 \end{array} \right) M \left( \begin{array}{c}
  \Phi_1^\dagger \Phi_1 \\
  \Phi_2^\dagger \Phi_2 \\
  \Phi_3^\dagger \Phi_3 \end{array} \right),
\ee
where $M$ is a real and symmetric matrix:
\be
\label{m}
M = \left( \begin{array}{ccc} a_1 & b_3 + X_3 & b_2 + X_2 \\
  b_3 + X_3 & a_2 & b_1 + X_1 \\ b_2 + X_2 & b_1 + X_1 & a_3
\end{array} \right).
\ee
In Eq.~\eqref{m},
\be
\label{xk}
X_k := c_k \varpi_k + d_k \Re \left( e^{i \varepsilon_k} \pi_k \right)
= \varpi_k
\left[ c_k + d_k \cos \left( \varepsilon_k + \vartheta_k \right) \right].
\ee
BnessFB means that $V_4$ is 
\textcolor{black}{positive}
for any values of the fields $a, \ldots, f$.
Since in Eq.~\eqref{mvo} 
\textcolor{black}{$\Phi_k^\dagger \Phi_k > 0\ \forall k$,}
this is equivalent to the matrix $M$ being
\textcolor{black}{strictly} copositive~\cite{kannike}.
The conditions for \textcolor{black}{strict} copositiveness
are\footnote{Since $Q \textcolor{black}{>} 0$,
condition~\eqref{vmp} is equivalent to either $P \ge 0$
or $P^2 \textcolor{black}{<} Q^2$;
the latter condition
is equivalent to $\det M \textcolor{black}{>} 0$.}
\bs
\label{dop}
\ba
a_1 &\textcolor{black}{>}& 0, \label{ur1} \\
a_2 &\textcolor{black}{>}& 0, \label{ur2} \\
a_3 &\textcolor{black}{>}& 0, \label{ur3} \\
B_1 \equiv b_1 + \sqrt{a_2 a_3} + X_1 &\textcolor{black}{>}& 0, \label{ur4} \\
B_2 \equiv b_2 + \sqrt{a_3 a_1} + X_2 &\textcolor{black}{>}& 0, \label{ur5} \\
B_3 \equiv b_3 + \sqrt{a_1 a_2} + X_3 &\textcolor{black}{>}& 0, \label{ur6} \\
P + Q &\textcolor{black}{>}& 0, \label{ur7}
\label{vmp}
\ea
\es
where
\bs
\ba
P &\equiv& \sqrt{\prod_{k=1}^3 a_k}
+ \sum_{k=1}^3 \left( b_k + X_k \right) \sqrt{a_k},
\\
Q &\equiv& \sqrt{2 B_1 B_2 B_3}.
\ea
\es
BnessFB is equivalent to Eqs.~\eqref{dop} holding
for all possible values of the $\pi_k$.
Using Eq.~\eqref{xk},
conditions~\eqref{ur4}--\eqref{ur6} may be written as
\be
\label{dfe}
\varpi_k
\left[ c_k + d_k \cos \left( \varepsilon_k + \vartheta_k \right) \right]
\textcolor{black}{>} - \bar b_k,
\ee
where
\be
\bar b_1 := b_1 + \sqrt{a_2 a_3}, \qquad
\bar b_2 := b_2 + \sqrt{a_3 a_1}, \qquad
\bar b_3 := b_3 + \sqrt{a_1 a_2}.
\ee
Since each $\varpi_k$ may take any value between $0$ and $1$,
and since each $\vartheta_k$ may take any value between $0$ and $2 \pi$,
condition~\eqref{dfe} is equivalent to $0 \textcolor{black}{>} - \bar b_k$
and $c_k - \left| d_k \right| \textcolor{black}{>} - \bar b_k$.
So,
altogether conditions~\eqref{ur1}--\eqref{ur6} are equivalent to
\textcolor{black}{the ``necessary conditions of level~1'' (NCL1),
\textit{viz.}}
\be
\label{2hdm}
\mathbf{NCL1}: \qquad
a_k \textcolor{black}{>} 0, \qquad \bar b_k \textcolor{black}{>} 0, \qquad
\bar b_k + e_k \textcolor{black}{>} 0
\qquad \forall k = 1, 2, 3,
\ee
where
\be
e_k := c_k - \left| d_k \right|.
\ee
Only condition~\eqref{ur7},
\textit{viz.}
\be
\label{jdp}
\sum_{k=1}^3 X_k \sqrt{a_k}
+ \sqrt{2 \prod_{k=1}^3 \left( \bar b_k + X_k \right)}
\textcolor{black}{>} - \bar a,
\ee
remains to be taken care of.
In inequality~\eqref{jdp},
\be
\bar a := \sqrt{ \prod_{k=1}^3 a_k} + \sum_{k=1}^3 b_k \sqrt{a_k}.
\ee
%
\textcolor{black}{The NCL1, \textit{i.e.}\ conditions~\eqref{2hdm},
  are necessary conditions
for BnessFB.}
They are well known:
if for instance the fields $e$ and $f$ are zero,
\textit{i.e.}\ if the doublet $\Phi_3$ is zero,
then one has the 2HDM formed by only $\Phi_1$ and $\Phi_2$
and with symmetry $\mathbb{Z}_2^{(2)}$;
the conditions $a_1 \textcolor{black}{>} 0$,
$a_2 \textcolor{black}{>} 0$,
$\bar b_3 \textcolor{black}{>} 0$,
and $\bar b_3 + e_3 \textcolor{black}{>} 0$
are precisely the BFB conditions for that 2HDM~\cite{sher}.
Thus,
the NCL1 may be obtained
by considering the three 2HDMs that are sub-cases of the 3HDM.

\paragraph{NCL2:}
In order to go beyond conditions~\eqref{2hdm}
one must use inequality~\eqref{jdp}.
One possibility is enforcing that inequality at the points $P_k$.
Using Eq.~\eqref{xk} one sees that,
for instance at point $P_1$ where $\varpi_2 = \varpi_3 = 0$ and $\varpi_1 = 1$,
$X_2 = X_3 = 0$ and the minimum possible value of $X_1$ is $e_1$.
One thus obtains for $P_1$,
$P_2$,
and $P_3$,
\be
\label{nc2}
\mathbf{NCL2}: \qquad \left\{ \begin{array}{rcl}
e_1 \sqrt{a_1} + \sqrt{2 \left( \bar b_1 + e_1 \right) \bar b_2 \bar b_3}
&\textcolor{black}{>} - \bar a,
\\*[1mm]
e_2 \sqrt{a_2} + \sqrt{2 \bar b_1 \left( \bar b_2 + e_2 \right) \bar b_3}
&\textcolor{black}{>} - \bar a,
\\*[1mm]
e_3 \sqrt{a_3} + \sqrt{2 \bar b_1 \bar b_2 \left( \bar b_3 + e_3 \right)}
&\textcolor{black}{>} - \bar a,
\end{array} \right.
\ee
respectively.
Inequalities~\eqref{nc2} are necessary conditions for BnessFB;
we call them the ``necessary conditions of level~2 (NCL2).''

\paragraph{NCL3:} Let us fix $\vartheta = 0$.
Condition~\eqref{jdp} then reads
\ba
\varpi_1 \left[ c_1
  + d_1 \cos \left( \varepsilon - \vartheta_2 - \vartheta_3 \right)
  \right] \sqrt{a_1}
+ \sum_{j=2}^3 \varpi_j \left( c_j + d_j \cos{\vartheta_j} \right) \sqrt{a_j}
& & \no
+ \sqrt{2 \left\{ \bar b_1 + \varpi_1 \left[ c_1 + d_1
    \cos \left( \varepsilon - \vartheta_2 - \vartheta_3 \right) \right]
  \right\}
  \prod_{j=2}^3 \left[ \bar b_j
    + \varpi_j \left( c_j + d_j \cos{\vartheta_j} \right) \right]}
&\textcolor{black}{>}& - \bar a,
\label{mfw}
\ea
where we have opted for the choice $\varepsilon_1 = \varepsilon$
and $\varepsilon_2 = \varepsilon_3 = 0$.
Inequality~\eqref{mfw}
must hold for all $\vartheta_2, \vartheta_3 \in \left[ 0, 2 \pi \right[$
and for all $\varpi_k$ that satisfy Eq.~\eqref{mbp} with $\vartheta = 0$,
\textit{viz.}\ with
\be
\varpi_1 = \left[ \sqrt{\varpi_2 \varpi_3} \pm
  \sqrt{\left( 1 - \varpi_2 \right) \left( 1 - \varpi_3 \right)} \right]^2.
\ee
So,
there are four degrees of freedom $\vartheta_2$,
$\vartheta_3$,
$\varpi_2$,
and $\varpi_3$ in inequality~\eqref{mfw}.
One may reduce them to two degrees of freedom in,
for instance,
one of the following two ways:
\begin{enumerate}
  \item One may consider point $P_4$ which,
    besides having $\vartheta = 0$,
    also has $\varpi_1 = \varpi_2 = \varpi_3 = 1$.
    Inequality~\eqref{mfw} then reads
    \ba
    \left[ c_1 + d_1 \cos
      \left( \varepsilon - \vartheta_2 - \vartheta_3 \right) \right] \sqrt{a_1}
    + \sum_{j=2}^3 \left( c_j + d_j \cos{\vartheta_j} \right) \sqrt{a_j}
    & & \no
    + \sqrt{2 \left[ \bar b_1 + c_1 + d_1 \cos
        \left( \varepsilon - \vartheta_2 - \vartheta_3 \right) \right]
      \prod_{j=2}^3 \left( \bar b_j + c_j + d_j \cos{\vartheta_j} \right)}
    &\textcolor{black}{>}& - \bar a.
    \label{mee}
    \ea
    Inequality~\eqref{mee} must hold for all values of the phases $\vartheta_2$
    and $\vartheta_3$.
    In particular,
    one may
    choose both $\vartheta_2$ and $\vartheta_3$
    to take one of the four values $\left\{ 0, \pi/2, \pi, 3\pi/2 \right\}$;
    one then obtains
    \be
    \mathbf{NCL3}: \qquad \sum_{k=1}^3 \,\beth_k\, \sqrt{a_k}
    + \sqrt{2 \prod_{k=1}^3 \left( \bar b_k + \beth_k \right)}
    \textcolor{black}{>} - \bar a,
    \label{ncl3}
    \ee
    with each of the following seven options for the $\beth_k$:
    \bs
    \label{nc3}
    \ba
    & & \beth_1 = c_1 + d_1 \cos{\varepsilon}, \qquad
    \beth_2 = c_2 + d_2, \qquad \beth_3 = c_3 + d_3;
    \\*[1mm]
    & &\beth_1 = c_1 + d_1 \cos{\varepsilon}, \qquad
    \beth_2 = c_2 - d_2, \qquad
    \beth_3 = c_3 - d_3;
    \\*[1mm]
    & & \beth_1 = c_1 - d_1 \cos{\varepsilon}, \qquad
    \beth_2 = c_2 - d_2, \qquad
    \beth_3 = c_3 + d_3;
    \\*[1mm]
    & & \beth_1 = c_1 - d_1 \cos{\varepsilon}, \qquad
    \beth_2 = c_2 + d_2, \qquad
    \beth_3 = c_3 - d_3;
    \\*[1mm]
    & & \beth_1 = c_1 - \left| d_1 \cos{\varepsilon} \right|, \qquad
    \beth_2 = c_2, \qquad
    \beth_3 = c_3;
    \\*[1mm]
    & & \beth_1 = c_1 - \left| d_1 \sin{\varepsilon} \right|, \qquad
    \beth_2 = c_2, \qquad
    \beth_3 = e_3.
    \\*[1mm]
    & & \beth_1 = c_1 - \left| d_1 \sin{\varepsilon} \right|, \qquad
    \beth_2 = e_2, \qquad
    \beth_3 = c_3.
    \ea
    \es
    Since we might instead have opted for either of the choices
    $\varepsilon_2 = \varepsilon$ and $\varepsilon_1 = \varepsilon_3 = 0$,
    or $\varepsilon_3 = \varepsilon$ and $\varepsilon_1 = \varepsilon_2 = 0$,
    one must also consider all the possibilities~\eqref{nc3}
    after permutation of the indices $1$,
    $2$,
    and $3$.
    Our ``necessary conditions of level~3 (NCL3)'' for BnessFB
    are inequality~\eqref{ncl3}
    with each of the seven choices~\eqref{nc3},
    together with their counterparts after permutation of the indices $1$,
    $2$,
    and $3$,
    making a total of 21 inequalities.
    \item One may choose both $\vartheta_2$ and $\vartheta_3$
    to take one of the four values $\left\{ 0, \pi/2, \pi, 3\pi/2 \right\}$
    while keeping both $\varpi_2$ and $\varpi_3$ free.
    Inequality~\eqref{mfw} then reads
    \ba
    \left[ \sqrt{\varpi_2 \varpi_3} \pm
      \sqrt{\left( 1 - \varpi_2 \right) \left( 1 - \varpi_3 \right)} \right]^2
    \,\beth_1\, \sqrt{a_1}
    + \sum_{j=2}^3 \varpi_j \,\beth_j \,\sqrt{a_j}
    & & \no
    + \sqrt{2 \left\{ \bar b_1 + \left[ \sqrt{\varpi_2 \varpi_3} \pm
      \sqrt{\left( 1 - \varpi_2 \right) \left( 1 - \varpi_3 \right)} \right]^2
      \beth_1 \right\}
      \prod_{j=2}^3 \left( \bar b_j + \varpi_j\, \beth_j \right)}
    &\textcolor{black}{>}& - \bar a,
    \label{err}
    \ea
    with each of the seven choices for the $\beth_k$
    displayed in Eqs.~\eqref{nc3}.
\end{enumerate}

\paragraph{NCL4:}
One may choose the alternative route
of using inequality~\eqref{mee} keeping $\vartheta_2$ and $\vartheta_3$ free,
together with inequality~\eqref{err}
with free $\varpi_2$ and $\varpi_3$.
Our ``necessary conditions of level~4 (NCL4)'' for BnessFB
consist of scanning inequality~\eqref{mee}
over both $\vartheta_2 \in \left[ 0, 2 \pi \right[$
and $\vartheta_3 \in \left[ 0, 2 \pi \right[$,
    together with scanning inequality~\eqref{err}---with both signs
    ``$+$'' and ``$-$'',
    and with all the 21 possibilities for the $\beth_k$ in Eqs.~\eqref{nc3}
and their counterparts with interchanged indices 1, 2, and 3---over
both $\varpi_2 \in \left] 0, 1 \right]$
and $\varpi_3 \in \left] 0, 1 \right]$.
In order to achieve greater accuracy,
one may additionally scan the inequality~\eqref{mfw}
by fixing $\vartheta_2=\pi$ and varying $\vartheta_3 \in [0,2\pi[$
    (and, conversely,
    fixing $\vartheta_3=\pi$
    and varying $\vartheta_2 \in \left[ 0, 2\pi \right[$),
        while simultaneously scanning $\varpi_{2,3} \in \left] 0, 1 \right]$.
    The same procedure may
    be repeated for the corresponding configurations
    obtained by permuting the indices 1, 2, and 3.

It should be emphasized that the inequalities in NCL2--NCL4
were selected on the basis of a painstaking study
of the minimization of the left-hand side of Eq.~\eqref{jdp}. 
Namely,
the left-hand side of Eq.~\eqref{jdp} tends to attain minimal values 
for certain regions of the parameter space
spanned by the $\varpi_k$ and $\vartheta_k$.
The inequalities in NCL2--NCL4 are the ones
with strongest discriminating power for testing BnessFB.

\section{$CP$-conserving case}
\label{sec:Branco's}

We refer to the $CP$-conserving
$\mathbb{Z}_2^{(3)} \times \mathbb{Z}_2^{(2)}$-symmetric $V_4$
as Branco model.
In that $V_4$,
the phases $\varepsilon_k$ satisfy $\varepsilon = 0$. 
Consequently,
the necessary BFB conditions NCL1--NCL4 of section~\ref{sec:necCond}
apply with $\varepsilon = 0$.
In particular,
the NCL1 and NCL2 do not depend on the $\varepsilon_k$.
The NCL3 are given by Eq.~\eqref{ncl3},
with $\beth_k$ chosen according to each of the following seven options:
    \bs
    \label{Branc_nc3}
    \ba
    \beth_1 = c_1 + d_1, & \beth_2 = c_2 + d_2, & \beth_3 = c_3 + d_3;
    \\*[1mm]
    \beth_1 = c_1 + d_1, & \beth_2 = c_2 - d_2, & \beth_3 = c_3 - d_3;
    \\*[1mm]
    \beth_1 = c_1 - d_1, & \beth_2 = c_2 - d_2, & \beth_3 = c_3 + d_3;
    \\*[1mm]
    \beth_1 = c_1 - d_1, & \beth_2 = c_2 + d_2, & \beth_3 = c_3 - d_3;
    \\*[1mm]
    \beth_1 = e_1, & \beth_2 = c_2, & \beth_3 = c_3;
    \\*[1mm]
    \beth_1 = c_1, & \beth_2 = e_2, & \beth_3 = c_3;
    \\*[1mm]
    \beth_1 = c_1, & \beth_2 = c_2, & \beth_3 = e_3.
    \ea
    \es
The NCL4 are implemented by scanning inequality
    \ba
    \left[ c_1 + d_1 \cos
      \left(\vartheta_2 + \vartheta_3 \right) \right] \sqrt{a_1}
    + \sum_{j=2}^3 \left( c_j + d_j \cos{\vartheta_j} \right) \sqrt{a_j}
    & & \no
    + \sqrt{2 \left[ \bar b_1 + c_1 + d_1 \cos
        \left(\vartheta_2 + \vartheta_3 \right) \right]
      \prod_{j=2}^3 \left( \bar b_j + c_j + d_j \cos{\vartheta_j} \right)}
    &\textcolor{black}{>}& - \bar a
    \label{Branc_mee}
    \ea
over both $\vartheta_2 \in \left[ 0, 2 \pi \right[$
and $\vartheta_3 \in \left[ 0, 2 \pi \right[$,
in combination with scanning inequality~\eqref{err}
over $\varpi_2 \in \left] 0, 1 \right]$
and $\varpi_3 \in \left] 0, 1 \right]$,
for each sign choice and with the $\beth_k$
defined in each of Eqs.~\eqref{Branc_nc3}.

\section{Sufficient conditions}
\label{sec:sufficient}

According to Ref.~\cite{GOO},
sufficient conditions for BnessFB of the quartic potential~\eqref{uty} are
\bs
\label{suffGOO_1}
\ba
a_k &\textcolor{black}{>}& 0, \\
f_1 &\textcolor{black}{>}& - \sqrt{a_2 a_3}, \\
f_2 &\textcolor{black}{>}& - \sqrt{a_3 a_1}, \\
f_3 &\textcolor{black}{>}& - \sqrt{a_1 a_2},
\ea
\es
and either
\be
\label{suffGOO_2}
\sum_{k=1}^3 f_k \sqrt{a_k} \textcolor{black}{>} 0
\ee
or
\be
\label{suffGOO_3}
\prod_{k=1}^3 a_k + 2 \prod_{k=1}^3 f_k - \sum_{k=1}^3 a_k f_k^2
\textcolor{black}{>} 0,
\ee
where
\be
\label{suffGOO_4}
f_k := b_k + \min \left( 0, e_k\right).
\ee

According to Ref.~\cite{boto},
sufficient conditions for BnessFB of the quartic potential~\eqref{uty} are
\bs
\label{suffBoto_1}
\ba
a_k &\textcolor{black}{>}& 0, \\
\bar g_k &\textcolor{black}{>}& 0, \\
\sqrt{\prod_{k=1}^3 a_k}
+ \sum_{k=1}^3 g_k \sqrt{a_k}
+ \sqrt{2 \prod_{k=1}^3 \bar g_k} &\textcolor{black}{>}& 0.
\ea
\es
Here,
\be
\label{suffBoto_2}
g_k := b_k + \min \left( 0, c_k\right) - \left| d_k \right|,
\ee
and
\be
\label{suffBoto_3}
\bar g_1 := g_1 + \sqrt{a_2 a_3}, \qquad 
\bar g_2 := g_2 + \sqrt{a_3 a_1}, \qquad 
\bar g_3 := g_3 + \sqrt{a_1 a_2}. 
\ee

Note that none of these two sets of sufficient conditions
involves the parameters $\varepsilon$;
they coincide for Weinberg model and Branco model.

We have numerically confirmed that
each of these two sets of sufficient conditions
selects a subset of the true BFB region
but excludes regions that are also BFB.
Moreover,
neither of the two subsets
selected by conditions~\eqref{suffGOO_1}--\eqref{suffGOO_3}
and~\eqref{suffBoto_1} is a subset of the other subset.
Last but not least,
we have found that the conditions~\eqref{suffBoto_1}
exclude a substantially larger portion of parameter space
than conditions~\eqref{suffGOO_1}–\eqref{suffGOO_3}.

\section{Minimization of the potential}
\label{sec:minimization}

Our analysis required the global minimization of $V_4$ for several purposes:
\begin{itemize}
\item In order to confirm that the quartic potentials
  that satisfy the sufficient conditions for BnessFB~\cite{GOO,boto}
  really are BFB.
\item In order to confirm that all the BFB quartic potentials
  do satisfy the necessary conditions for BnessFB.
  \item In order to quantify the accuracy of the necessary conditions,
  \textit{i.e.}\ to find out the percentage of the potentials
  that obey those conditions which are \emph{not} BFB.
\item In order to prepare training data for our NNs.
\end{itemize}
For the minimization we have used the \Math function {\tt NMinimize}.
Since that function operates on real variables,
the complex scalar fields were decomposed
into their real and imaginary components. 
We fixed $a = 0$ and $b = 1$
and we minimized $V_4$ relative to the remaining fields.
Fixing $a=0$ is always possible via a local $SU(2)\times U(1)$
transformation and allows one to reduce the number of real variables by two.
Fixing $b=1$ further reduces the number of real variables by one
and improves the stability of the numerical minimization,
without entailing any lack of generality, since it simply corresponds
to a re-scaling of all the nonzero fields $b$, $c$, $d$, $e$, and $f$.
Note that the particular case $a=b=0$ requires no separate treatment
as it corresponds to the 2HDM
formed by the doublets $\Phi_2$ and $\Phi_3$, for
which the necessary and sufficient BnessFB conditions are NCL1.

Global minimization is computationally very demanding and may
return a local minimum instead of the global one.
In order to improve robustness and accuracy,
we employed the following strategies:
\begin{itemize}
\item Minimization with multiple algorithms,
  including Nelder--Mead and Differential Evolution.\footnote{Nelder--Mead
  is a derivative-free optimization method
  that searches the parameter space by using a simplex.
  It is not a true global optimization algorithm;
  nevertheless,
  in practice it often performs well for problems
  with a relatively modest number of local minima.
  Differential Evolution is a population-based stochastic method
  designed for global optimization.
  Although computationally expensive,
  it is relatively robust and tends to work well
  for problems that have several local minima.}
\item Minimization both with the default numerical precision
  and with increased precision.
\item Minimization over multiple field ranges,
  \textit{i.e.}\ allowing the real and imaginary parts of $c$,
  $d$,
  $e$,
  and $f$ to vary from $-m$ to $+m$
  for various values of the positive number $m$.
\item In Weinberg model,
  minimization of $V_4$ by using the three equivalent representations
  given in Eqs.~\eqref{three_forms}.
\end{itemize}
For each $V_4$,
\textit{i.e.}\ for each parameter set
\be
\left\{ a_1, a_2, a_3, b_1, b_2, b_3, c_1, c_2, c_3, d_1, d_2, d_3, \epsilon
\right\},
\label{set}
\ee
the minimization was repeated multiple times using the procedures above
with different random seed points. 
If any run ever produced a negative value of $V_4$,
then the minimization cycle would be interrupted,
since that negative value definitely means that that $V_4$ is not BFB.
Because Nelder--Mead is relatively fast,
it was applied more extensively,
in particular over a broader set of field ranges,
\textit{i.e.}\ for $m = 10^5, 10^4, 700, 200, 20, 1$. 
Differential Evolution usually is
more robust,
\textit{i.e.}\ it usually produces a smaller percentage of incorrect results,
but it is substantially more expensive;
therefore,
it was applied only for
$m = 10^5, 10^4, 500$.

Repeating the minimization with different algorithms,
precisions,
and initial conditions avoids getting trapped at local minima
with steep gradients. 
Cross-checks of the minimization results indicate that
the adopted procedure is highly accurate,
albeit computationally quite intensive. 
In order to ensure reliability
and to assess the accuracy of the necessary BFB conditions,
we have performed minimization for many data sets,
each of them containing one or more millions of quartic potentials.

\section{\SW}
\label{sec:package}

\subsection{Setup}
\label{sec:setup}

\paragraph{Installation:}
\SW\ is publicly available at the site
\begin{description}
\item \url{https://github.com/jurciukonis/StableWein}.
\end{description}
The package may either be installed by cloning the repository
\begin{lstlisting}[language=bash]
git clone https://github.com/jurciukonis/StableWein.git
\end{lstlisting}
or downloaded directly from the
\href{https://github.com/jurciukonis/StableWein}{GitHub repository}.
Once downloaded,
the archive should be extracted and the resulting folder
(named {\tt StableWein})
may be placed in a directory chosen by the user. 
\SW\ contains the following components:
\begin{itemize}
\item {\tt StableWein/StableWein.nb}: the main package file;
\item {\tt StableWein/Binaries/}:
  binary files generated for compiled functions;
\item {\tt StableWein/Documentation/}: documentation files;
\item {\tt StableWein/Nets/}: trained neural networks.
\end{itemize}
The package may be loaded by evaluating \SW\ from any \Math session through the command
\begin{lstlisting}[language=Mathematica,morekeywords={NotebookEvaluate,FileNameJoin}]
NotebookEvaluate[FileNameJoin[{NotebookDirectory[EvaluationNotebook[]], "WB_package/WB_package.nb"}]]
\end{lstlisting}
by providing the appropriate path to {\tt StableWein.nb}.

\paragraph{Compatibility:}
We verified that \SW\ correctly
operates with \Math versions from 13.2 up to the latest release,
\textit{viz.}\ 14.3.
The package was also tested on Linux,
macOS,
and Windows operating systems. 

\SW\ makes use of the {\tt Compile} function to generate compiled code.
In order to enable the option {\tt CompilationTarget -> "C"},
a suitable external C compiler must be installed. 
Some versions of \Math (\textit{viz.}\ the 14.x releases)
generate compiled functions that execute substantially slower
than those produced by earlier versions. 
In particular,
we have observed such performance problems
for compiled code containing iterative constructs such as {\tt Do},
{\tt For},
{\tt Sum} and related loop-based operations within {\tt Compile}.
In order to mitigate this problem,
we provide the option to load pre-compiled functions
generated by using \Math 13.2.
In order to enable this feature
the global boolean variable {\tt wbLoadBinaries}
must be set to {\tt True} prior to loading the package,
\emph{i.e.}\ {\tt wbLoadBinaries = True}.

To improve computational efficiency,
we employ global functions instead of local functions.
All our global functions are named with a prefix {\tt "wb"}
so that they do not conflict with any other global functions.
The same prefix is used in the names of our global variables.

\subsection{Package structure}
\label{sec:structure}

The structure of the package,
together with a brief description of the corresponding functionality,
is summarized below:
\begin{itemize}
\item[\textbf{+}] {\tt wbWeinUniBFBcheck}:
  main function for
  Weinberg model.
    \begin{itemize}
    \item[+] {\blue UNI}:
      implementation of the unitarity constraints.
        \begin{itemize}
        \item[+] {\green Generation}:
          generation of quartic-potential couplings.
            \begin{itemize}
            \item[--] {\red UNI}:
              filtering of the generated couplings
              through
              the UNI constraints.
            \item[--] {\red UNI + NCL1}:
              filtering of the generated couplings
              through both the
              UNI and NCL1 constraints.
            \end{itemize}
          \item[+] {\green Filtration}:
            filtering of an imported data set.
            \begin{itemize}
            \item[--] {\red UNI}:
              filtering of couplings
              through
              the UNI constraints.
            \end{itemize}            
        \end{itemize}
      \item[+] {\blue NNs}:
        neural network utilities.
        \begin{itemize}
        \item[--] {\green Predictions}:
          prediction of couplings satisfying both the UNI and BFB conditions.
        \item[--] {\green Predictions + validation}:
          subsequent validation of the predicted data sets
          by using the selected BFB method(s).
        \end{itemize}        
      \item[+] {\blue BFB}:
        implementation of the BFB constraints.
        \begin{itemize}
        \item[--] {\green Sufficient conditions (mode~0)}:
          application of the sufficient BFB conditions.
        \item[+] {\green Necessary conditions}:
          application of the necessary BFB conditions.
            \begin{itemize}
            \item[--] {\red Mode~1}:
              application of the NCL1.
            \item[--] {\red Mode~2}:
              application of the NCL1, NCL2, and NCL3.
            \item[--] {\red Mode~3}:
              application of the NCL1, NCL2, NCL3, and NCL4.
            \end{itemize}
          \item[--] {\green Mode~4}:
           application of brute force to minimize $V_4$ and find out whether it can ever be negative.
        \end{itemize}
    \end{itemize}
  \item[\textbf{+}] {\tt wbBrancUniBFBcheck}: main function
    for Branco model,
    with the same sub-items
    as for Weinberg model,
    as listed above.  
\end{itemize}

As is evident from the package structure described above,
the package has two main entry points:
{\tt wbWeinUniBFBcheck},
which selects couplings satisfying the UNI\footnote{
These refer to the perturbative unitarity bounds
derived in Appendix~\ref{app:unitarity}.}
and BFB constraints for Weinberg model,
and {\tt wbBrancUniBFBcheck},
which does the same task for Branco model. 
The internal structure of these two routines is identical. 
Each of them comprises three main components:
(i) verification of the unitarity conditions,
(ii) verification of the bounded-from-below conditions,
(iii) prediction of viable couplings by using neural networks.

The package may either generate new sets
of couplings\footnote{The generation is done
by using a uniform distribution
for each coupling within its pre-assigned range.
Users
who dislike this option
must generate their sets of couplings by other means.}
or filter imported sets by applying user-selected constraints
via the options listed in Table~\ref{tab:options}. 
\begin{table}[!h]
\begin{center}
\begin{tabular}
{@{\hspace{2mm}}
>{\raggedright\arraybackslash}p{3.3cm}
>{}p{3.3cm}
>{}p{8.5cm}
@{\hspace{2mm}}}
\hlinewd{1.1pt}
option & default value & meaning \\
\hline\\[-1.5mm]
        {\tt GeneratePoints} &
        {\tt True} &
        generates the initial data set \\[2.5mm]
        {\tt NPoints} & $10^3$ &
        number of initial points in the data set \\[2.5mm]
        {\tt MaxCoupling} & 16 &
        upper bound for each generated coupling \\[2.5mm]
\hline\\[-1.5mm]
        {\tt ImposeUNI} &
        {\tt True} &
        filters the couplings through the UNI conditions \\[2.5mm]
        {\tt BoundUNI} & $8 \pi$ &
        unitarity bound (see Appendix~\ref{app:unitarity}) \\[2.5mm]        
\hline\\[-1.5mm]
        {\tt ImposeBFB} &
        {\tt True} &
        filters the couplings
        through the BFB conditions defined by set value of mode \\[2.5mm]
\hline\\[-1.5mm]
        {\tt NeuralNets} & {\tt False} &
        uses neural networks \\[2.5mm]
\hline\\[-1.5mm]
        {\tt ExportData} & {\tt False} &
        exports the initial data set and the validated data set
        to a file \\[2.5mm]
        {\tt FileNameInit} & {\tt initial_set.csv} &
        file name for the initial data set \\[2.5mm]
        {\tt FileNameValid} & {\tt valid_set.csv} &
        file name for the validated data set \\[2.5mm]        
\hline\\[-1.5mm]
        {\tt ShowMessages} & {\tt True} &
        displays messages reporting the computation times and results \\[2.5mm]
\hlinewd{1.1pt}
\end{tabular}
\end{center}
\vspace{-3mm}
\caption{Options for the functions {\tt wbBrancUniBFBcheck}
  and {\tt wbWeinUniBFBcheck}.}
\label{tab:options}
\end{table}
%
In the generation mode,
the couplings are sampled randomly within prescribed ranges
and subsequently filtered to satisfy the chosen UNI and/or BFB requirements.
Alternatively,
neural networks may be used to predict couplings;
in this case,
the networks predict couplings
that satisfy the UNI and BFB constraints simultaneously;
the predicted sets of couplings may afterwards,
if one so wishes,
be validated
by using the selected BFB procedures.

The accuracy of the BFB-based filtering
is controlled by the choice of mode = 0\,$\ldots$\,4
in the {\tt wbWeinUniBFBcheck} and {\tt wbBrancUniBFBcheck} functions.
These functions support a hierarchy of increasingly stringent BFB tests,
which trade computational cost for completeness.
Mode~0 corresponds to the sufficient BFB conditions; 
all the higher modes begin by applying those conditions. 
This strategy optimizes performance,
because quartic potentials that satisfy the sufficient conditions
are immediately classified as being BFB;
additional checks are required only for potentials
that fail the sufficiency test.

In the lowest-accuracy modes,
the BFB filter enforces
smaller sets of
necessary analytic constraints.
Higher-accuracy modes incorporate additional necessary conditions,
progressively reducing the allowed parameter region. 
In mode~3,
the numerical evaluation relies on a discretized sampling
of the relevant orbit-space surface; 
the sampling resolution may be increased to improve accuracy,
at the expense of longer running times.
In the most stringent mode~4,
BnessFB is checked through brute-force numerical
minimization of the quartic potential.
This provides an explicit cross-check of the
necessary conditions
implemented in modes 1--3,
but with a much longer running time.

Although the package is designed such that users
interact mainly with the 
functions {\tt wbWeinUniBFBcheck} and {\tt wbBrancUniBFBcheck},
all the submodules may also be executed independently
by calling the corresponding functions listed in Table~\ref{tab:functions}.
\begin{table}[!h]
\begin{normalsize}
\normalsize
\begin{center}
\begin{tabular}
{@{\hspace{2mm}}
>{\raggedright\arraybackslash}p{4.0cm}
>{}p{11.7cm}
@{\hspace{2mm}}}
\hlinewd{1.1pt}
function & action \\
\hline\\[-1.5mm]
        {\tt wbWeinParRnd, wbWeinParRndNec,
          wbWeinParRndUni, wbWeinParRndUniNec} &
        generates random, uniformly distributed,
        couplings of $V_4$ in the format~\eqref{set}
        within different ranges \\[2.5mm]
\hline\\[-1.5mm]
        {\tt wbWeinUNICond} &
        filters the couplings through
        the perturbative unitarity conditions \\[2.5mm]
\hline\\[-1.5mm]
        {\tt wbWeinUNINec1Cond} &
        identical to {\tt wbWeinParRndUni},
        but includes additional filtering through the NCL1 \\[2.5mm]
\hline\\[-1.5mm]
        {\tt wbWeinSufCond} &
        filters the couplings through the sufficient BFB conditions \\[2.5mm]
\hline\\[-1.5mm]
        {\tt wbWeinNecCond1} &
        filters the couplings through the NCL1  \\[2.5mm]
        {\tt wbWeinNecCond2} &
        filters the couplings through the NCL1, NCL2, and NCL3  \\[2.5mm]
        {\tt wbWeinNecCond3} &
        filters the couplings through the NCL1, NCL2, NCL3, and NCL4
        \\[2.5mm]
\hline\\[-1.5mm]
          {\tt wbMinimFuncPotWein} &
          minimizes $V_4$ \\[2.5mm]
          {\tt wbWeinGridGen} &
          auxiliary function for generation of the scanning grid \\[2.5mm]
          {\tt wbWeinUniBFBcheck} &
          returns couplings satisfying the UNI and BFB constraints
          with configurable method and accuracy \\[2.5mm]
\hlinewd{1.1pt}
\end{tabular}
\end{center}
\vspace{-3mm}
\end{normalsize}
\caption{Functions for Weinberg model.
  For Branco model
  the functions are analogous and their names have the same structure,
  but with the substring {\tt Wein} replaced by {\tt Branc}.
  }
\label{tab:functions}
\end{table}

\subsection{Usage}
\label{sec:usage}

We illustrate the use of the package
with the function {\tt wbWeinUniBFBcheck};
the function {\tt wbBrancUniBFBcheck} is used analogously.
Each of those functions supports two calling patterns:
\begin{itemize}
\item {\tt wbWeinUniBFBcheck[mode, options]},
\item {\tt wbWeinUniBFBcheck[list, mode, options]}. 
\end{itemize}
The second pattern uses a data set furnished by the user
instead of generating one.
Note that the order or the couplings
is the one specified in Eq.~\eqref{set};
in Branco model,
the generated couplings take the form
$\left\{ a_1, a_2, a_3, b_1, b_2, b_3, c_1, c_2, c_3, d_1, d_2, d_3 \right\}$.
In both cases,
the output is a pair of data sets,
where the first one is the initial or generated data set
and the second one is the filtered or validated data set.
A simple example of using the function would be the command
\begin{lstlisting}[language=Mathematica,morekeywords={wbWeinUniBFBcheck}]
{initParamSet, validParamSet} = 
  wbWeinUniBFBcheck[3, "NPoints" -> 10^4, "ExportData" -> True];
\end{lstlisting}
which generates $10^4$ sets of couplings satisfying the UNI conditions,
then filters them through BFB mode~3,
and exports the validated data sets to a file.
Note that {\tt wbWeinUniBFBcheck}
does not export the data to files unless instructed to do so;
since a variety of data formats may be used,
the users are allowed
to perform import and export operations
in \Math as needed. 
The imported data may
be provided to {\tt wbWeinUniBFBcheck}
by setting the global parameter {\tt wbParamData}:
\begin{lstlisting}[language=Mathematica,morekeywords={wbWeinUniBFBcheck,FileNameJoin}]
initFile = 
  Import[FileNameJoin[NotebookDirectory[] <> "/Data/initial_set.csv"]];
wbParamData = initFile;
{initParamSet, validParamSet} = wbWeinUniBFBcheck[3, "GeneratePoints" -> False];
\end{lstlisting}
Alternatively,
the imported data may be passed directly
to {\tt wbWeinUniBFBcheck}---in this case,
the option {\tt "GeneratePoints" -> False} may be omitted:
\begin{lstlisting}[language=Mathematica,morekeywords={wbWeinUniBFBcheck}]
{initParamSet, validParamSet} = wbWeinUniBFBcheck[initFile, 3];
\end{lstlisting}

By default,
{\tt wbWeinUniBFBcheck} generates an initial data set
with the chosen constraints imposed.
The number of generated points is controlled by the option {\tt NPoints}. 
By setting {\tt "ImposeUNI" -> True|False}
and {\tt "ImposeBFB" -> True|False},
one may generate an initial data set that either does or does not satisfy
the perturbative unitarity and/or the BnessFB NCL1. 
The unitarity constraints are
described in Appendix~\ref{app:unitarity};
the moduli of the eigenvalues of the scattering matrices
are bounded via the option {\tt BoundUNI},
which is set to $8\pi$ by default;
if a different unitarity bound is wished for,
or if only the BFB constraints are desired,
then the range of the generated couplings
should be specified by using the option {\tt MaxCoupling}. 
In particular,
a relaxed unitarity bound may be used,
e.g.\
\begin{lstlisting}[language=Mathematica,morekeywords={wbWeinUniBFBcheck}]
{initParamSet,validParamSet} = wbWeinUniBFBcheck[1,"NPoints" -> 10^4, "MaxCoupling" -> 35, "BoundUNI" -> 16*Pi, "ImposeBFB" -> False];
\end{lstlisting}
Through this command,
$10^4$ potentials satisfying
the UNI conditions with the unitarity bound $16\pi$ 
and with each coupling smaller than 35 in modulus are generated,
and no BFB conditions are enforced.

The routine also supports generating viable data sets
by using neural network predictions;
this is done by setting {\tt "NeuralNets" -> True}. 
The predictors are trained to propose couplings
that are expected to satisfy the unitarity and BFB requirements. 
If {\tt "ImposeBFB" -> True} is also set,
then the predicted points are additionally validated
by using the selected BFB method (modes 1--4);
the UNI check is already incorporated into the neural network generation step.

Mode~3 relies on scanning grids that approximate the orbit-space surface
represented in Fig.~\ref{fig:allowed}. 
The default grid settings are chosen to be computationally efficient
while remaining highly accurate for most practical applications
(approximately $99.99\%$ accuracy).
The default grid parameters for {\tt wbWeinUniBFBcheck} are
{\tt stepsWein = \{17, 40, 0\}},
whereas for the function {\tt wbBrancUniBFBcheck}
the default grid is {\tt stepsBranc = \{17, 15\}}.\footnote{The list
{\tt stepsWein = \{ p, r, s \}},
where $p$,
$r$,
and $s$ are non-negative integers,
determines the grid sizes.
The integer $p$ specifies the grid used for scanning over the $\varpi_k$.
The finest nominal step is $1/p$;
however,
the scan includes all the rational values $n/m$
with $1 \le m \le p$ and $1 \le n \le m$.
The integer $r$ defines the grid for the angles $\vartheta_k$,
with finest resolution $\pi/r$.
The integer $s$ controls an additional scan associated with
Eq.~\eqref{mfw},
wherein $\vartheta_2$ is varied with fixed $\vartheta_3 = \pi$ and vice
versa.
In order to make a finest scan one must increase $p$ and/or $r$ and/or
$s$.
The orbit-space surface is sampled in a randomized manner
while maintaining an approximately uniform
and duplicate-free distribution of points.
The grids for the $\varpi_k$ and $\vartheta_k$
are computed and stored in memory when the package is loaded;
this grid preparation significantly improves
the computational efficiency of mode~3 as compared with a direct scan.
In the case of Branco model,
only two grid parameters are required,
\textit{viz.}\ {\tt \{ p, r \}}.}
In order to further improve the accuracy
it may be necessary to increase the grid resolution 
by raising the integer parameters in {\tt stepsWein} and {\tt stepsBranc} 
and then applying the function {\tt wbWeinGridGen}:
\begin{lstlisting}[language=Mathematica,morekeywords={wbWeinGridGen}]
stepsWein = {27, 80, 30};
{wbfaceListWein, wb\[Theta]ListWein, wb\[Theta]ListWein2}
= wbWeinGridGen[stepsWein];
\end{lstlisting}
We warn the user that increasing the grid density
may substantially raise the running time of mode~3 filtering
while producing only minor changes
in the number of accepted potentials.

Maximal accuracy is achieved through mode~4,
where quartic potentials in the initial data set are validated
through their direct minimization,
in accordance with Sec.~\ref{sec:minimization}.
This mode is useful for checking the BFB-validated points
obtained in modes 1--3. 
We warn the user that the minimization of quartic potentials
is computationally expensive and requires substantial computational resources.

The status of intermediate computations
is displayed at the bottom of the evaluation notebook. 
In addition,
messages reporting computation times and results
are printed at the end of each step;
by setting {\tt "ShowMessages" -> False},
these messages are suppressed; 
this is convenient when {\tt wbWeinUniBFBcheck}
is used inside loops or similar workflows.
For instance,
\begin{lstlisting}[language=Mathematica,morekeywords={wbWeinUniBFBcheck}]
Table[mode -> Length[Last[wbWeinUniBFBcheck[initParamSet,mode,"GeneratePoints" -> False,"ShowMessages" -> False]]], {mode,1,4}]
\end{lstlisting}

We urge the user to check examples in documentation files,
which might be the simplest way to get acquainted with the package.
The usage messages and documentation files
may be accessed by typing a question mark before any function,
\textit{viz.}~{\tt ?wbWeinUniBFBcheck}.

\subsection{Examples}
\label{sec:examples}

The computations reported in this subsection
included four options:
both
\begin{description}
\item computations performed for Weinberg model,
reported in Table~\ref{tab:WeinRes2};
\item computations performed for Branco model,
reported in Table~\ref{tab:BrancRes2};
\end{description}
and also
\begin{description}
\item computations performed on a desktop computer
equipped with an Intel Core i9-13900K CPU (24 cores)
and 128~GB of RAM;
\item computations performed on an Apple MacBook Pro laptop
equipped with an M4 Pro CPU (14 cores) and 48~GB of RAM.
\end{description}
All configurations were evaluated under identical conditions,
with computations executed in parallel across all available CPU cores.
The computations performed on the Intel Core i9 used \Math 13.2.
Computations performed on the M4 Pro CPU used \Math 14.1;
in this case and in order to avoid the performance difficulties
discussed in Sec~\ref{sec:setup},
pre-compiled binaries were used by setting {\tt wbLoadBinaries = True}.

In each one of the four options,
we have generated a data set containing $10^6$ quartic potentials
that satisfy both the unitarity conditions
and the BnessFB NCL1.\footnote{The initial data sets
were generated by using the pseudo-random uniform distribution
implemented in {\tt wbWeinParRndUniNec} and {\tt wbBrancParRndUniNec}.
The percentages quoted in Tables~\ref{tab:WeinRes2}
and~\ref{tab:BrancRes2} might differ if different distributions
were used to generate the initial points.}
The data sets were then further refined
by using both {\tt wbWeinUniBFBcheck} and {\tt wbBrancUniBFBcheck}
in the following modes:
\begin{itemize}
\item Mode~0: the sufficient BnessFB conditions
  of Refs.~\cite{GOO} and~\cite{boto} were applied.
\item Mode~2: the necessary BnessFB conditions NCL2 and NCL3,
  which require no scanning,
  were applied.
\item Mode~3: besides NCL2 and NCL3,
  the necessary BnessFB condition NCL4,
  with the default scanning grids of \SW,
  was applied.
\item Mode~3 +: besides NCL2 and NCL3,
  the necessary BnessFB condition NCL4,
  with the ameliorated scanning grid +,
  which is $\{ 17, 80, 0 \}$ for Weinberg model
  and $\{ 40, 30 \}$ for Branco model,
  was used.
\item Mode 4: the potential was minimized by brute force.
\end{itemize}
\begin{figure}[h!]
 \centering
	\includegraphics[width=0.9\textwidth]{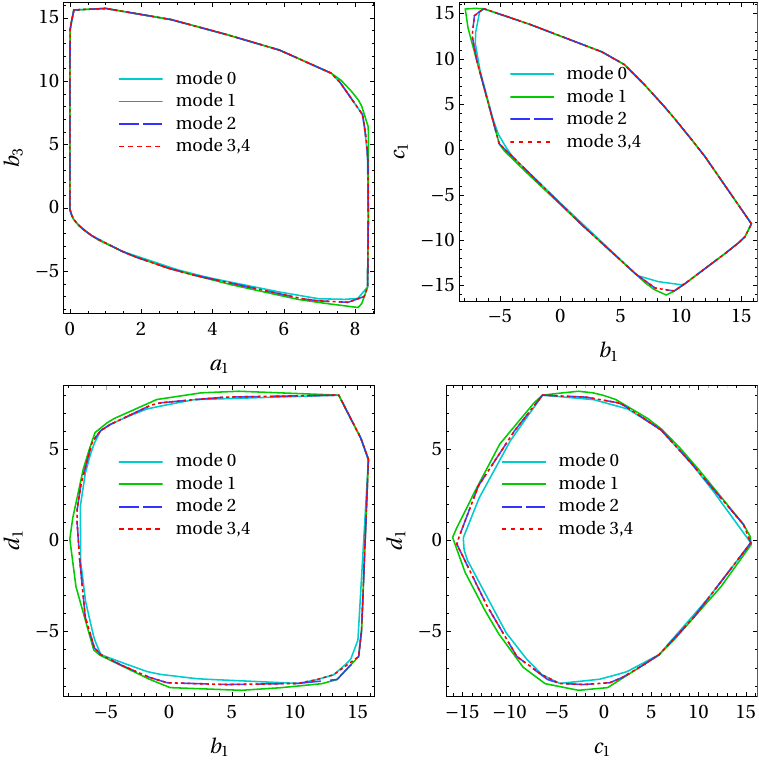}
  \caption{Regions of the parameter space of the Weinberg model
    that both satisfy the unitarity conditions
    and pass the BFB constraints at different accuracy modes.}
  \label{fig:regions}
\end{figure}
The accuracy of modes 2 and 3 was assessed
by comparing their BFB-accepted potentials
with those obtained from brute-force minimization,
\textit{viz.}\ mode~4,
according to the formula
\begin{equation}
     \mathrm{Accuracy} \left( \mathrm{\tt mode\ } i \right) =
     \frac{\mathrm{BFB} \left( \mathrm{\tt mode\ } 4 \right)}
     {\mathrm{BFB} \left( \mathrm{\tt mode\ } i \right)} \times 100\%,
\end{equation}
where $\mathrm{BFB} \left( \mathrm{\tt mode\ } i \right)$
is the number of BFB potentials according to $\mathrm{\tt mode\ } i$
for $i = 2, 3, 4$.

\begin{table}[h!]
\begin{center}
\begin{tabular}
{@{\hspace{2.mm}}
>{\raggedright\arraybackslash}p{2.4cm}|
p{2.1cm}
p{1.7cm}
p{1.6cm}|
p{2.1cm}
p{1.7cm}
p{1.6cm}
@{\hspace{1.7mm}}}
\hlinewd{1.1pt}
\empty & \multicolumn{3}{c|}{Intel Core i9} & \multicolumn{3}{c}{Apple M4 Pro} \\
 & BFB points & Accuracy  & Time (s) & BFB points & Accuracy  & Time (s) \\
\hline
&&&&&&\\[-2mm]
    {\tt Generation}
    & 
$1\,000\,000$ & --- & $382$ &   
$1\,000\,000$ & --- & $533$\\ [2.5mm]
{\tt mode 0} & 
$602\,536$ & --- & $11$ &  
$600\,847$ & --- & $4$\\ [2.5mm]
{\tt mode 2} & 
$754\,471$ & $99.360\%$ & $16$ &  
$753\,115$ & $99.361\%$ & $7$\\ [2.5mm]
{\tt mode 3} & 
$749\,674$ & $99.996\%$ & $112$ &  
$748\,343$ & $99.995\%$ & $559$\\ [2.5mm]
{\tt mode 3 +}
&
$749\,666$ & $99.997\%$ & $693$  &
$748\,328$ & $99.997\%$ & $3747$\\ [2.5mm]
{\tt mode 4} &
$749\,640$ & $100\%$ &  $31\,360$ &
$748\,302$ & $100\%$ & $47\,961$\\ [2.5mm]
\hlinewd{1.1pt}
\end{tabular}
\end{center}
\vspace{-3mm}
\caption{Comparison of the results obtained
  with different computational modes using {\tt wbWeinUniBFBcheck}.
}
\label{tab:WeinRes2}
\end{table}

\begin{table}[h!]
\begin{center}
\begin{tabular}
{@{\hspace{2.mm}}
>{\raggedright\arraybackslash}p{2.4cm}|
p{2.1cm}
p{1.7cm}
p{1.6cm}|
p{2.1cm}
p{1.7cm}
p{1.6cm}
@{\hspace{1.7mm}}}
\hlinewd{1.1pt}
\empty & \multicolumn{3}{c|}{Intel Core i9} & \multicolumn{3}{c}{Apple M4 Pro} \\
 & BFB points & Accuracy  & Time (s) & BFB points & Accuracy  & Time (s) \\
\hline
&&&&&&\\[-2mm]
    {\tt Generation}
    & 
$1\,000\,000$ & --- & $386$ &   
$1\,000\,000$ & --- & $529$\\ [2.5mm]
{\tt mode 0} & 
$601\,644$ & --- & $10$ &  
$602\,566$ & --- & $4$\\ [2.5mm]
{\tt mode 2} & 
$750\,018$ & $99.987\%$ & $14$ &  
$751\,171$ & $99.990\%$ & $8$\\ [2.5mm]
{\tt mode 3} & 
$749\,925$ & $99.999\%$ & $15$ &  
$751\,093$ & $100\%$ & $40$\\ [2.5mm]
{\tt mode 3 +}
&
$749\,923$ & $100\%$ & $107$  &
--- & --- & ---\\ [2.5mm]
{\tt mode 4} &
$749\,923$ & $100\%$ &  $23\,941$ &
$751\,093$ & $100\%$ & $36\,260$\\ [2.5mm]
\hlinewd{1.1pt}
\end{tabular}
\end{center}
\vspace{-3mm}
\caption{Comparison of the results obtained
  with different computational modes using {\tt wbBrancUniBFBcheck}.
}
\label{tab:BrancRes2}
\end{table}

Figure~\ref{fig:regions} compares
the boundaries of the allowed regions
obtained for different modes of the BFB
constraints, as projected on different two-dimensional planes
of parameter space.
We find that,
although the sufficient conditions
reject nearly 20\% of the BFB-allowed points,
the corresponding contours in parameter space
are only slightly smaller than those obtained from the full BFB test. 
By contrast,
the BFB NCL1 yield slightly larger contours.

\subsection{Neural networks}
\label{sec:neuralNets}

The neural networks were trained to predict the UNI and BFB constraints
simultaneously. 
Because the UNI conditions restrict the allowed ranges
of the parameters,
the effective training domain is reduced,
which improves the attainable predictive accuracy. 
The neural networks were constructed, trained, and deployed using the capabilities of the Wolfram Language.
We employ fully connected feed-forward neural networks with eight layers. 
In addition,
we consider four architectures that differ only in their width,
with $2^n$ neurons per layer for $n = 7,8,9,10$; 
these architectures are denoted by $\mathbb{N}_7, \ldots, \mathbb{N}_{10}$,
respectively. 
The number of layers and neurons was chosen to optimize
the trade-off between accuracy and computational efficiency. 
Based on computational experiments that we have performed,
the best performance is obtained with the {\tt Ramp} activation function
and Xavier weight initialization~\cite{Xavier}.

We found that augmenting the training with synthetic data
improves the predictive accuracy.
Accordingly,
we trained the networks on multiple data sets
with different class proportions in order to increase diversity.
The first training set was constructed from false samples
drawn from the raw data
(generated with {\tt wbWeinParRndUniNec} and {\tt wbWeinUNICond})
and supplemented with 20\% true samples;
the latter were validated through direct minimization
of the potential.\footnote{Here,
a ``true sample'' denotes a potential
that satisfies both the UNI and BFB conditions,
whereas a ``false sample'' is a potential that violates
at least one of those conditions.} 
The two smallest networks,
$\mathbb{N}_7^{(1)}$ and $\mathbb{N}_8^{(1)}$,
were trained on this data set and subsequently used to generate
refined synthetic samples,
which were again validated by using the UNI and BFB criteria. 
Each of the first and second training sets
contained $2.5 \times 10^7$ samples. 
All four networks $\mathbb{N}_7^{(2)}$ through $\mathbb{N}_{10}^{(2)}$
were then trained on the second training set.

A third training set was constructed
by combining the synthetic samples with the refined,
newly generated samples; 
it contains $5 \times 10^7$ entries. 
All four networks $\mathbb{N}_7^{(3)}$ through $\mathbb{N}_{10}^{(3)}$
were subsequently trained on this third data set. 
Each network was trained for 500 epochs
with a batch size of $2^{14}$ samples,
by using the {\tt "RMSProp"} optimizer, 
which adapts the learning rate via an exponentially smoothed estimate
of the gradient magnitude.
Although large training data sets were used in our study,
smaller data sets may be used in practice,
since the predictive accuracy improves only marginally
when one increases the training-set size.
\begin{figure}[b!]
 \centering
	\includegraphics[width=\textwidth]{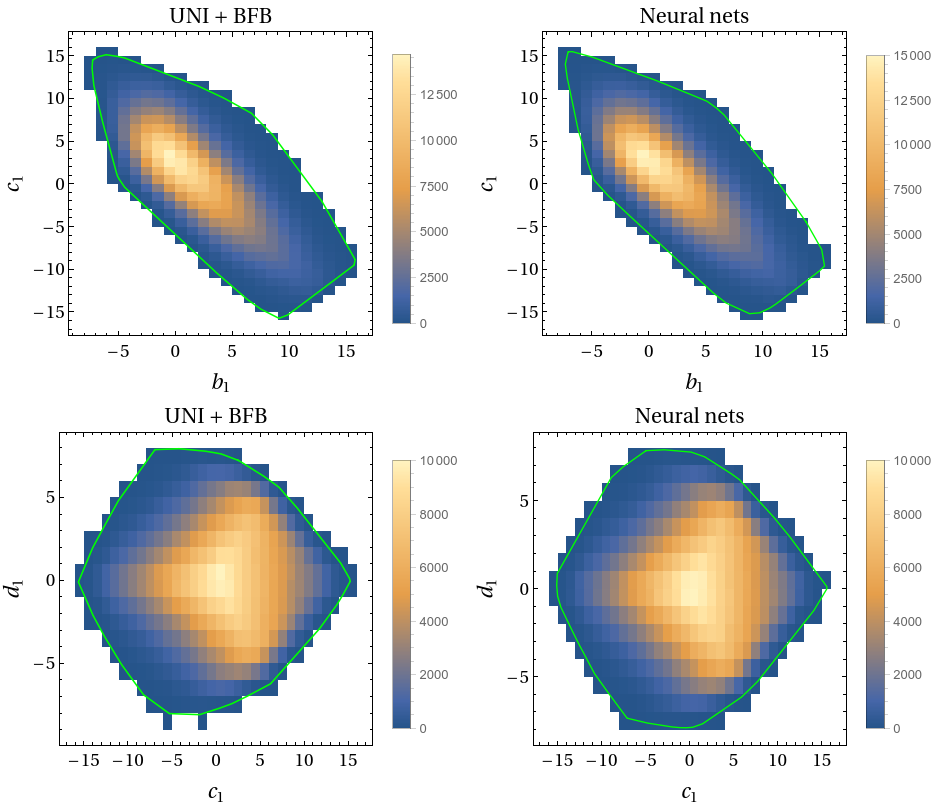}
  \caption{Density histograms of the parameter space
    in Weinberg model
    satisfying the UNI constraints and passing the BFB conditions.
    The left column shows data generated
    by using {\tt wbWeinUniBFBcheck} with mode~3,
    while the right column shows data generated by using neural networks.}
  \label{fig:net_distrib}
\end{figure}

We constructed the following neural network chain to classify the potentials:
\begin{equation}
  \mathbb{N}_7^{(1)}
  \rightarrow \mathbb{N}_7^{(2)}
  \rightarrow \mathbb{N}_7^{(3)}
  \rightarrow \mathbb{N}_8^{(1)}
  \rightarrow \mathbb{N}_8^{(2)}
  \rightarrow \mathbb{N}_8^{(3)}
  \rightarrow \mathbb{N}_9^{(2)}
  \rightarrow \mathbb{N}_9^{(3)}
  \rightarrow \mathbb{N}_{10}^{(2)}
  \rightarrow \mathbb{N}_{10}^{(3)}, 
\label{netChain}
\end{equation}
where each subsequent network reclassifies the points
accepted by its predecessor. 
The smaller architectures $\mathbb{N}_7$ and $\mathbb{N}_8$
offer faster inference but lower accuracy when applied directly to raw data,
whereas the larger architectures $\mathbb{N}_9$ and $\mathbb{N}_{10}$
achieve higher accuracy at the expense of increased computational cost. 
Accordingly,
combining light-weight and high-capacity networks
in the chain~\eqref{netChain}
yields high overall accuracy while maintaining efficient inference.

\SW\ can generate data using NNs by setting {\tt "NeuralNets" -> True}
in the functions {\tt wbWeinUniBFBcheck} and {\tt wbBrancUniBFBcheck}. 
If {\tt "ImposeBFB" -> True} is additionally specified,
then the predicted potentials are further validated
by using the selected BFB-accuracy mode. 
For example,
the following command generates a data set of one million points:
\begin{lstlisting}[language=Mathematica,morekeywords={wbWeinUniBFBcheck}]
{initParamSet, validParamSet} =
wbWeinUniBFBcheck[3, "NeuralNets" -> True, "NPoints" -> 10^6];
\end{lstlisting}
On a system equipped with an Intel Core i9 processor,
generating this data set requires 1\,340~s. 
Enforcing mode~3 yields a neural network prediction accuracy of 99.95\%. 
Although the inference time of the neural network
exceeds by one order of magnitude the total running time
required both for data generation and subsequent refinement with mode~3
(see Table~\ref{tab:WeinRes2}),
it still remains one order of magnitude shorter
than the time required for direct minimization of the potentials. 
We also recall that the BFB-filtering times
reported in Table~\ref{tab:WeinRes2}
correspond to processing only the $\sim 40\%$ of the points
that do not satisfy the sufficient conditions. 

One may try and adapt
the proposed machine learning technique
to provide an efficient tool for classifying BFB potentials
in models for which neither sufficient nor necessary BFB conditions are known,
\textit{e.g.}\ three-Higgs-doublet models
without $\mathbb{Z}_2 \times \mathbb{Z}_2$ symmetry.

Figure~\ref{fig:net_distrib} compares the density of couplings
generated by applying the UNI and BFB conditions directly (left panel)
to couplings generated by the neural network predictor (right panel).
The two two-dimensional projections of parameter space show
that the resulting distributions are similar.

It should be noted that on multi-socket systems,
\textit{e.g.}~workstations and servers,
neural network inference may be slower,
due to suboptimal thread scheduling and memory-management settings.
Such a slowdown of neural network inference
was observed by us on a system equipped with an M4 Pro chip.

\section{Conclusions}
\label{sec:conclusions}

In this work we have analyzed
conditions for
the scalar potential of the $\mathbb{Z}_2 \times \mathbb{Z}_2$-symmetric 
three-Higgs-doublet model
to be
bounded from below (BFB).
Rather than striving for
sufficient conditions
(which select only a portion of parameter space,
neglecting potentially physically interesting parts of that space), 
we argue that,
for this model,
it is possible to write ever more necessary conditions
that progressively approximate the boundary of the BFB region.

As explained in Sec.~\ref{sec:minimization},
our computational effort to reliably minimize
$\mathbb{Z}_2 \times \mathbb{Z}_2$-symmetric 3HDM potentials
has allowed us to understand better
the boundary separating BFB from non-BFB potentials in parameter space
(see, for instance, Fig.~\ref{fig:regions}).
By doing so,
we were able to compare the accuracy of various necessary conditions.
This pursuit culminated in the set of necessary conditions
that we call NCL1--4, which can be used to predict,
with very high (well above 99\%) accuracy,
whether a potential is BFB or not,
without undue computational effort.
Based on our results,
we produced and make available the \Math package \SW,
that allows users to predict with varying levels of accuracy
whether a $\mathbb{Z}_2 \times \mathbb{Z}_2$-symmetric
3HDM
potential is BFB,
by either generating or inputting a set of quartic couplings.

We have found that the two sets of sufficient conditions of 
Refs.~\cite{GOO,boto} guarantee that $\sim 60\%$ of randomly generated
potentials satisfying UNI and NCL1 are BFB.
Conditions~\eqref{suffBoto_1} guarantee that about $22\%$
of the potentials are BFB,
and conditions~\eqref{suffGOO_1}–\eqref{suffGOO_3} do the same
for $\sim 60\%$ of the potentials.
Still,
we have found that neither set of sufficient conditions
is a subset of the other one,
and we have implemented them both in \SW.

By comparing with numerically minimized sets of quartic potentials (mode 4), 
we have found that $\sim 75\%$ of the potentials satisfying NCL1 (mode 1)
are BFB;
moreover,
even the simple combination of necessary conditions
NCL1--3 (mode 2) achieves close to $99.9\%$ accuracy.
The NCL4 (mode 3) requires a scan over the surface of the parameter space
represented in Fig.~\ref{fig:allowed};
although computationally more expensive than modes 1 and 2,
we have found that NCL4 allow one to reach $\sim 99.999\%$ accuracy
in scanning potentials for BnessFB.
Tables~\ref{tab:WeinRes2} and~\ref{tab:BrancRes2} 
show that the brute-force numerical minimization used in mode~4
is slower than mode~3 by a factor of $\mathcal{O}(10^3)$
and slower than mode~2 by a factor of $\mathcal{O}(10^4)$,
revealing the disproportion between the computational time required for mode~4
and the very high accuracy attained by NCL1--4.

NCL1--4 are computationally less expensive
for Branco model.
For this reason, 
instead of treating this model
as the particular case of Weinberg model with $\varepsilon = 0$,
we have implemented specialized functions
to handle Branco 
model
efficiently.

Our necessary conditions are based on reducing the
problem of BnessFB to the \textcolor{black}{strict} copositivity
of the matrix in Eq.~\eqref{m}.
Consequently,
our methods apply only to 3HDM potentials that enjoy
$\mathbb{Z}_2 \times \mathbb{Z}_2$ symmetry.
The problem of BnessFB
of non-$\mathbb{Z}_2 \times \mathbb{Z}_2$-symmetric potentials
remains untouched.

In addition, the \SW\ package provides a neural network option that has been trained to identify BFB
$\mathbb{Z}_2^{(3)} \times \mathbb{Z}_2^{(2)}$-symmetric potentials.
This approach, which relies on neither sufficient nor necessary analytic BFB conditions, 
achieves good performance, with an accuracy close to 99.9\% and fast inference.
Consequently, the proposed machine-learning technique can be adapted to classify BFB parameter points in models for which neither sufficient nor necessary BFB conditions are currently available,
\textit{viz.}\ in 3HDM without $\mathbb{Z}_2 \times \mathbb{Z}_2$ symmetry.

\vspace*{5mm}

\paragraph{Acknowledgements:} L.L.\ thanks Rafael Boto and Anton Kun\v{c}inas
for extensive discussions.
D.J.\ thanks Art\={u}ras Acus for valuable discussions.
The work of D.J.\ received funding from the Research Council of Lithuania (LMT)
under Contract No.~S-CERN-24-2;
part of the computations were performed
using the infrastructure of the Lithuanian Particle Physics Consortium
in the framework of agreement No.~VS-13 of Vilnius University with LMT.
The work of L.L.\ and A.M.\ was supported
by the Portuguese Foundation for Science and Technology (FCT)
through projects UIDB/00777/2020 and UIDP/00777/2020,
and by the Recovery and Resilience Plan within the scope
of investment RE-C06-i06,
measure RE-C06-i06.m02,
project 2024.01362.CERN.
The work of A.M.\ was further supported by FCT with PhD Grant No.~2024.01340.BD.

\newpage

\begin{appendix}

\setcounter{equation}{0}
\renewcommand{\theequation}{A\arabic{equation}}

\section{The possible symmetries of a 3HDM}
\label{app:symmetries}

\textcolor{black}{
Table~\ref{table:symmetries} gives all the internal symmetries
that a three-Higgs-doublet model may have
\emph{in the scalar potential alone}.\footnote{We claim
neither that these symmetries may be extended to the fermion sector,
\textit{viz.}\ to the Yukawa interactions,
nor that the models thus obtained are realistic.}
This classification
has been systematically developed in Refs.~\cite{vdovin1,vdovin2,vdovin3}
(additional details are given in Ref.~\cite{nishi});
a few groups were missed in the classification
and were added later~\cite{vazao,kuncinas1,bree,kuncinas2,kuncinas3}.)
}
\begin{table}[t]
  {\caption{Possible symmetries of the scalar potential
      of a three-Higgs-doublet model.
      First column: symmetries with $U(1)_1 \times U(1)_2$ subgroup.
      Second column: symmetries without $U(1)_1 \times U(1)_2$ subgroup
      but with $\mathbb{Z}_2^{(3)} \times U(1)_2$ subgroup.
      Third column: symmetries with none of the above subgroups
      but with $\mathbb{Z}_2^{(3)} \times \mathbb{Z}_2^{(2)}$ subgroup.
      Fourth column: other symmetries.
      The symmetries in red produce a scalar potential
      that automatically enjoys invariance
      under the $CP$ symmetry $\mathbb{Z}_2^\ast$;
      the symmetries in blue lead to a potential
      that has both a $\mathbb{Z}_2^\ast$-invariant version
      and a $CP$-breaking one.
      (\textit{Nota bene}: The potential with both
      $\left[ \mathbb{Z}_2^{(3)} \times \mathbb{Z}_2^{(2)} \right]
      \rtimes \bar{\mathbb{Z}}_2^\ast$ and $\mathbb{Z}_2^\ast$ symmetries
      is the $D_4$-invariant one;
      the potential with both $A_4$ and $\mathbb{Z}_2^\ast$ symmetries
      is the $S_4$-invariant one.)
      The symmetry in green is generated by the `generalized $CP$'
      transformation~\eqref{z4*}.
    }
    \label{table:symmetries}}
    \renewcommand{\arraystretch}{1.2}
  \begin{center}
    \begin{tabular}{c|c|c|c}
      \hlinewd{1.1pt}&&&\\[-16pt]
      $\textcolor{red}{U(1)_1 \times U(1)_2}$ &
      $\textcolor{red}{\mathbb{Z}_2^{(3)} \times U(1)_2}$ &
      $\textcolor{blue}{\mathbb{Z}_2^{(3)} \times \mathbb{Z}_2^{(2)}}$ &
      $\textcolor{blue}{\mathbb{Z}_2^{(3)}}$
      \\
      $\textcolor{red}{\left[ U(1)_1 \times U(1)_2 \right] \rtimes S_3}$ &
      $\textcolor{red}{D_4 \times U(1)_2}$ &
      $\textcolor{blue}{
        \left[ \mathbb{Z}_2^{(3)} \times \mathbb{Z}_2^{(2)} \right]
        \rtimes \bar{\mathbb{Z}}_2^\ast}$ &
      $\textcolor{blue}{U(1)_2}$
      \\
      $\textcolor{red}{SU(3)}$ &
      $\textcolor{red}{O(2) \times U(1)_2}$ &
      $\textcolor{red}{D_4}$ &
      $\textcolor{red}{U(1)_1}$
      \\
      $\textcolor{red}{U(2)}$ &
      &
      $\textcolor{red}{O(2)}$ &
      $\textcolor{blue}{\mathbb{Z}_3}$
      \\
      &
      &
      $\textcolor{blue}{A_4}$ &
      $\textcolor{red}{\mathbb{Z}_4}$
      \\
      &
      &
      $\textcolor{red}{S_4}$ &
      $\textcolor{red}{\mathbb{Z}_2^\ast}$
      \\
      &
      &
      $\textcolor{red}{SO(3)}$ &
      $\textcolor{green}{\mathbb{Z}_4^\ast}$
      \\
      &
      &
      &
      $\textcolor{blue}{S_3^\prime}$
      \\
      &
      &
      &
      $\textcolor{blue}{\Delta (54)}$ \\
      &
      &
      &
      $\textcolor{red}{\Sigma (36)}$ \\[4pt]
      \hlinewd{1.1pt}
    \end{tabular}
  \end{center}
\end{table}

\textcolor{black}{The symmetries in Table~\ref{table:symmetries}
  are the `realizable' ones,
  \textit{i.e.} the symmetry groups of the potential
  after any accidental symmetries have been included.}

\textcolor{black}{In principle,
  we should have discarded from Table~\ref{table:symmetries}
  any subgroup of the $U(1)$ gauge group of hypercharge,
  because hypercharge is not an \emph{internal} symmetry.
  A case in point is the group $\mathbb{Z}_3^\mathrm{center}$,
  formed by the transformations~\eqref{u12} with $\varsigma = 0$,
  $2 \pi / 3$,
  and $4 \pi / 3$.
  That group is the center of the $SU(3)$ group that mixes the three doublets,
  which \emph{is} an internal symmetry,
  but it is also a subgroup of the $U(1)$ of hypercharge,
  which \emph{is not} an internal symmetry.
  Therefore,
  $U(1)_2$ should rather be written
  $\left. U(1)_2 \right/ \mathbb{Z}_3^\mathrm{center}$.
  On the one hand,
  for the sake of simplicity,
  in Table~\ref{table:symmetries}
  we have ommitted writing ``$\left/ \mathbb{Z}_3^\mathrm{center} \right.$''
  in all the groups containing $U(1)_2$,
  and also in $SU(3)$ and $\Delta (54)$.
  On the other hand,
  and also for the sake of simplicity,
  in Table~\ref{table:symmetries}
  we have used $\Sigma (36)$
  instead of $\Sigma (36 \times 3)$---the latter group
  includes $\mathbb{Z}_3^\mathrm{center}$,
  and $\Sigma (36) \cong
  \left. \Sigma (36 \times 3) \right/ \mathbb{Z}_3^\mathrm{center}$.
}

The symmetry $\mathbb{Z}_2^\ast$ is the usual $CP$ symmetry generated by
\be
\label{z2*}
\mathbb{Z}_2^\ast: \quad
\Phi_k \to  \Phi_k^\ast,\ \forall k = 1, 2, 3.
\ee
The symmetry $\bar{\mathbb{Z}}_2^\ast$ is the $CP$ symmetry generated by
\be
\label{barz2*}
\bar{\mathbb{Z}}_2^\ast: \quad
\Phi_1 \to  \Phi_1^\ast, \quad
\Phi_2 \to  \Phi_3^\ast, \quad
\Phi_3 \to  \Phi_2^\ast.
\ee
The symmetry $\mathbb{Z}_4^\ast$ is the $CP$ symmetry generated by
\be
\label{z4*}
\mathbb{Z}_4^\ast: \quad
\Phi_1 \to  \Phi_1^\ast, \quad
\Phi_2 \to  \Phi_3^\ast, \quad
\Phi_3 \to  - \Phi_2^\ast.
\ee
Many symmetries---the ones
in red in Table~\ref{table:symmetries}---automatically
lead to a scalar potential that is invariant under $\mathbb{Z}_2^\ast$;
other symmetries---the ones in blue---have both
a $\mathbb{Z}_2^\ast$-invariant version
and a $\mathbb{Z}_2^\ast$-noninvariant one.

The generators of $U(1)_1$ and $U(1)_2$ are given in Eqs.~\eqref{u1u1};
the generators of $\mathbb{Z}_2^{(3)}$ and $\mathbb{Z}_2^{(2)}$
are given in Eqs.~\eqref{z23} and~\eqref{z22},
respectively.
The symmetry $\mathbb{Z}_4$ is generated by the transformation
\be
\mathbb{Z}_4: \quad \Phi_2 \to i \Phi_2, \quad \Phi_3 \to - i \Phi_3,
\ee
which is a particular case of the $U(1)_1$ transformation~\eqref{u11}.
The symmetry $\mathbb{Z}_3$ is generated by the transformation
\be
\label{z3}
\mathbb{Z}_3: \quad
\Phi_2 \to e^{2 i \pi / 3} \Phi_2, \quad \Phi_3 \to e^{- 2 i \pi / 3} \Phi_3,
\ee
which is another particular case of the $U(1)_1$ transformation.
The symmetry $S_3^\prime$ in the last column of Table~\ref{table:symmetries}
has generators~\eqref{z3} and
\be
\label{z223}
\mathbb{Z}_2^{2 \leftrightarrow 3}: \quad \Phi_2 \leftrightarrow \Phi_3.
\ee
The symmetry $D_4$ is generated by the transformations~\eqref{z23}
and~\eqref{z223}.\footnote{The symmetry $D_4 \times U(1)_2$
should rather be called
$\left[ D_4 \times U(1)_2 \right] \left/ \left[
  \mathbb{Z}_3^\mathrm{center} \times \mathbb{Z}_2^{(2,3)} \right] \right.$,
since the transformation
\be
\mathbb{Z}_2^{(2,3)}: \quad \Phi_2 \to - \Phi_2,\ \Phi_3 \to - \Phi_3
\ee
belongs both to $D_4$ and to $U(1)_2$.
Reference~\cite{kuncinas3} argues that $D_4 \times U(1)_2$
should rather be called
$U(1) \circ \left( \mathbb{Z}_2 \times \mathbb{Z}_2 \right)$.
\label{foot:d4}}
The symmetry $\Delta (54)$ has generators~\eqref{z3},
\eqref{z223},
and
%
\be
\label{z212}
\mathbb{Z}_2^{1 \leftrightarrow 2}: \quad \Phi_1 \leftrightarrow \Phi_2.
\ee
The symmetry $S_3$ in the first column
of Table~\ref{table:symmetries} is generated by
the transformations~\eqref{z223} and~\eqref{z212}.
The symmetry $\Sigma (36)$ has generators~\eqref{z3},
\eqref{z212},
and
\be
\left( \begin{array}{ccc} \Phi_1 \\ \Phi_2 \\ \Phi_3 \end{array} \right)
\to
\frac{1}{\sqrt{3}}
\left( \begin{array}{ccc} 1 \qquad 1 \qquad 1 \\
  1 \qquad e^{2 i \pi / 3} \qquad e^{- 2 i \pi / 3} \\
  1 \qquad e^{- 2 i \pi / 3} \qquad e^{2 i \pi / 3}
\end{array} \right)
\left( \begin{array}{ccc} \Phi_1 \\ \Phi_2 \\ \Phi_3 \end{array} \right).
\ee

The symmetry $A_4$ is generated by
\bs
\label{a4}
\ba
\mathbb{Z}_2^{(1,2)}: \qquad \qquad \Phi_1 \to - \Phi_1,\ \Phi_2 \to - \Phi_2;
\\
\mathbb{Z}_3^\mathrm{cyclical}: \qquad \qquad
\Phi_1 \to \Phi_2 \to \Phi_3 \to \Phi_1. 
\ea
\es
The symmetry $S_4$
is generated by the transformations~\eqref{z212} and~\eqref{a4}.

The symmetry $SO(3)$ is generated by the transformations
\bs
\label{so3}
\ba
\left( \begin{array}{c} \Phi_1 \\ \Phi_2 \end{array} \right)
&\to &
\left( \begin{array}{cc} \cos{\varepsilon} \qquad - \sin{\varepsilon} \\
\sin{\varepsilon} \qquad \cos{\varepsilon} \end{array} \right)
\left( \begin{array}{c} \Phi_1 \\ \Phi_2 \end{array} \right),
\label{12r}
\\
\left( \begin{array}{c} \Phi_2 \\ \Phi_3 \end{array} \right)
&\to &
\left( \begin{array}{cc} \cos{\varkappa} \qquad - \sin{\varkappa} \\
\sin{\varkappa} \qquad \cos{\varkappa} \end{array} \right)
\left( \begin{array}{c} \Phi_2 \\ \Phi_3 \end{array} \right),
\label{23r}
\ea
\es
with arbitrary angles $\varepsilon$ and $\varkappa$.
The symmetry $SU(3)$ is generated by the transformations~\eqref{so3}
and~\eqref{u1u1}.
The symmetry $O(2)$ is generated by the transformations~\eqref{z23}
and~\eqref{23r}.
The symmetry $U(2)$ is generated by the transformations~\eqref{u11}
and~\eqref{23r}.

\textcolor{black}{The $A_4$-symmetric 3HDM
  has the quartic potential of Eq.~\eqref{uty} with
  \be
  a_1 = a_2 = a_3, \quad b_1 = b_2 = b_3, \quad
  c_1 = c_2 = c_3, \quad d_1 = d_2 = d_3.
  \label{jgp}
  \ee
  The $S_4$-symmetric 3HDM is $A_4$-symmetric
  and besides has $e^{i \epsilon} = \pm 1$.
  The $SO(3)$-symmetric 3HDM is $S_4$-symmetric and besides has
  \be
  \label{ubo}
  a_2 = b_1 + c_1 + d_1.
  \ee
}

\textcolor{black}{The $D_4$-symmetric 3HDM also has $e^{i \epsilon} = \pm 1$
  and moreover
  \be
  \label{jco}
  a_2 = a_3, \quad b_2 = b_3, \quad c_2 = c_3, \quad d_2 = d_3.
  \ee
  The $O(2)$-symmetric 3HDM is $D_4$-symmetric
  and obeys Eq.~\eqref{ubo}.
  The $\left[ \mathbb{Z}_2^{(3)} \times \mathbb{Z}_2^{(2)} \right]
  \rtimes \bar{\mathbb{Z}}_2^\ast$-symmetric potential
  obeys Eqs.~\eqref{jco} but has free $\varepsilon$.
}

\newpage

\setcounter{equation}{0}
\renewcommand{\theequation}{B\arabic{equation}}

\section{Perturbative unitarity bounds (UNI)}
\label{app:unitarity}

In addition to satisfying the BnessFB conditions,
the couplings of the quartic part of the potential
must remain in a regime where perturbation theory is valid.
Only then does the $S$-matrix admit a controlled expansion
in powers of the interaction couplings,
thus preserving unitarity.
A particular class of theoretical constraints
enforcing this requirement arises
from the partial-wave decomposition of $2 \to 2$ scattering amplitudes.
These constraints are referred to as partial-wave unitarity bounds.

Consider the $2 \to 2$ scattering process of scalar fields $AB \to CD$,
with a Lorentz invariant amplitude $\mathcal{M}\left[ AB \to CD \right]$.
In the center-of-mass frame,
we can perform a partial-wave expansion
by decomposing the amplitude
in the basis of the Legendre polynomials $P_J(x)$ as
\begin{equation}
	\mathcal{M}\left[ AB \to CD \right] = 
	16\pi \sum_{J=0}^{\infty} \,a_J
        \left( 2J+1 \right)
        P_J(\cos\theta),
\end{equation}
where $\theta$ is the scattering angle,
$J = 0, 1, 2, \ldots$
is the total angular momentum of the initial/final state,
and the numerical coefficients $a_J$ are the partial waves. 
The orthogonality relations among the $P_J(x)$
allow the partial waves to be determined as
\begin{equation}
\label{ajdef}
	a_J = \frac{1}{32\pi} \int_{-1}^{+1} \mathrm{d}(\cos\theta)\,
	\mathcal{M}\left[ AB \to CD \right]
	P_J(\cos\theta) .
\end{equation}
The
unitarity of the $S$-matrix (more specifically
the optical theorem)
implies
that the partial waves must reside within the disk defined by~\cite{Hally}
\begin{equation}
\label{opt_th}
\Im a_J
\geq |a_J|^2,\ \forall J,
\end{equation}
with equality holding for purely elastic processes.
The unitarity constraint~\eqref{opt_th} may be rewritten as
\begin{equation}
\label{IMRE}
\left|a_J\right| \leq  1,\qquad
0 \leq
\Im a_J
\leq 1, \qquad
\left|
\Re a_J
\right| \leq \frac{1}{2},\quad
\forall J.
\end{equation}

Let us assume that each partial wave
admits an expansion in powers of
a coupling $\lambda$:
\begin{equation}
\label{aJ_expand}
a_J =
\sum_{n=1}^\infty \lambda^n a_J^{(n)}.
\end{equation}
While
each full non-perturbative partial wave $a_J$
must satisfy conditions~\eqref{IMRE},
a truncated perturbative calculation need not.
However,
if the partial wave unitarity constraints are not satisfied at tree-level,
then higher-order corrections cannot
be
subdominant,
which means that the perturbative expansion is
unwarranted.
Since tree-level amplitudes are strictly real,
it is a necessary condition for the validity of perturbation theory
that the tree-level partial waves satisfy
\begin{equation}
\label{theUNI}
\left| \Re a_J^{(1)} \right|
\leq \frac{1}{2}.
\end{equation}

At tree-level,
a $2\to2$ scattering process may have contributions from $s$,
$t$,
and $u$-channel exchange diagrams,
as well as from contact interactions.
However,
in the ultra-relativistic limit and away from resonances,
the Mandelstam variable $s$ becomes large,
and then the tree-level amplitude is dominated by the quartic interactions
contained in $V_4$.
In this limit, the scattering amplitude reads
\begin{eqnarray}
\label{Msinf}
	\lim_{s\to \infty}\mathcal{M}\left[ AB \to CD \right] =
	-N_{AB}
        N_{CD}\,
        \frac{\partial^4 V_4}{\partial A \,\partial B \,
          \partial C^* \,\partial D^*},
\end{eqnarray}
where $N_{ij} = 2^{-\delta_{ij}/2}$ is a symmetry factor
that accounts for identical particles
in either the initial state or the final state.
Since expression~\eqref{Msinf} is independent of the scattering angle $\theta$,
and since $P_0(\cos\theta)=1$,
the strongest perturbative unitarity constraint
arises from the zeroth partial wave.
Using Eqs.~\eqref{ajdef} and \eqref{Msinf},
the necessary condition~\eqref{theUNI} becomes
\begin{equation}
\label{PWUC}
\left|\Re a_0^{(1)} \right|
\leq \frac{1}{2} 
	\quad\Leftrightarrow\quad
	\left|N_{AB}
        N_{CD}\,
        \frac{\partial^4 V_4}{\partial A \,\partial B \,
          \partial C^* \,\partial D^*}\right| \leq 8\pi.
\end{equation}

Condition~\eqref{PWUC} is basis-independent,
meaning it must be satisfied for \emph{any}
linear superposition of initial/final two-particle states.
The states in the superposition must have
the same quantum numbers under the symmetries of the theory.
It is therefore convenient to organize all
the states with two scalar fields
into coupled-channel scattering matrices,
whose entries are given by the corresponding zeroth partial waves
in the high-energy limit.
For each set of states with the same quantum numbers,
perturbative unitarity requires
that the \emph{moduli} of the \emph{eigenvalues}
of the coupled-channel matrix
remain below $8\pi$.

For the $\mathbb{Z}_2^{(3)} \times \mathbb{Z}_2^{(2)}$-symmetric 3HDM,
we label each two-particle state according to
\begin{equation}
  \left| Q,\, Y,\, T,\, \mathbb{Z}_2^{(2)},\,
  \mathbb{Z}_2^{(3)} \right\rangle,
\end{equation}
where $Q$ is the total electric charge,
$Y$ the total hypercharge,
and $T$ the total isospin.
As expected,
some eigenvalues are identical.
The set of two-particle states and the corresponding
coupled-channel matrices leading to the minimal set
of independent eigenvalues is given by~\cite{Lopes}
\begin{itemize}
\item $| 2,1,1,+,+ \rangle$: the possible two-particle states are 
  $\left\{ aa,\, cc,\, ee \right\}$,
  and the corresponding scattering matrix reads
		\begin{equation}
		\label{M33_1}
			\begin{pmatrix}
				a_1 & d_3 & d_2 \\
				d_3 & a_2 & d_1e^{i\varepsilon} \\
				d_2 & d_1e^{-i\varepsilon} & a_3 \\
			\end{pmatrix}.
		\end{equation}	
\item $| 2,1,1,+,- \rangle$: the only two-particle state is 
  $\left\{ ae \right\}$,
  and the corresponding scattering matrix reads
		\begin{equation}
			\begin{pmatrix}
				b_2 + c_2 \\
			\end{pmatrix}.
		\end{equation}	
\item $| 2,1,1,-,+ \rangle$: the only two-particle state is 
  $\left\{ ac \right\}$,
  and the corresponding scattering matrix reads
		\begin{equation}
			\begin{pmatrix}
				b_3 + c_3 \\
			\end{pmatrix}.
		\end{equation}	
\item $| 2,1,1,-,- \rangle$: the only two-particle state is 
  $\left\{ ce \right\}$,
  and the corresponding scattering matrix reads
		\begin{equation}
			\begin{pmatrix}
				b_1 + c_1 \\
			\end{pmatrix}.
		\end{equation}	
\item $| 1,1,0,+,- \rangle$: the only two-particle state is 
  $\left\{ \left( af-be \right) \left/ \sqrt{2} \right. \right\}$,
  and the corresponding scattering matrix reads
		\begin{equation}
			\begin{pmatrix}
				b_2 - c_2 \\
			\end{pmatrix}.
		\end{equation}	
\item $| 1,1,0,-,+ \rangle$: the only two-particle state is 
  $\left\{ \left( ad-bc \right) \left/ \sqrt{2} \right. \right\}$,
  and the corresponding scattering matrix reads
		\begin{equation}
			\begin{pmatrix}
				b_3 - c_3 \\
			\end{pmatrix}.
		\end{equation}	
\item $| 1,1,0,-,- \rangle$: the only two-particle state is 
  $\left\{ \left( cf-de \right) \left/ \sqrt{2} \right. \right\}$,
  and the corresponding scattering matrix reads
		\begin{equation}
			\begin{pmatrix}
				b_1 - c_1 \\
			\end{pmatrix}.
		\end{equation}	
\item $| 1,0,1,+,+ \rangle$: the possible two-particle states are 
  $\left\{ ab^*,\, cd^*,\, ef^* \right\}$,
  and the corresponding scattering matrix reads
		\begin{equation}
		\label{M33_2}
			\begin{pmatrix}
				a_1 & c_3 & c_2 \\
				c_3 & a_2 & c_1 \\
				c_2 & c_1 & a_3 \\
			\end{pmatrix}.
		\end{equation}	
\item $| 1,0,1,+,- \rangle$: the possible two-particle states are 
  $\left\{ eb^*,\, af^* \right\}$,
  and the corresponding scattering matrix reads
		\begin{equation}
			\begin{pmatrix}
				b_2 & d_2 \\
				d_2 & b_2 \\
			\end{pmatrix}.
		\end{equation}	
\item $| 1,0,1,-,+ \rangle$: the possible two-particle states are 
  $\left\{ cb^*,\, ad^* \right\}$,
  and the corresponding scattering matrix reads
		\begin{equation}
			\begin{pmatrix}
				b_3 & d_3 \\
				d_3 & b_3 \\
			\end{pmatrix}.
		\end{equation}	
\item $| 1,0,1,-,- \rangle$: the possible two-particle states are 
  $\left\{ ed^*,\, cf^* \right\}$,
  and the corresponding scattering matrix reads
		\begin{equation}
			\begin{pmatrix}
				b_1 & d_1e^{-i\varepsilon} \\
				d_1e^{i\varepsilon} & b_1 \\
			\end{pmatrix}.
		\end{equation}	
              \item $| 0,0,0,+,+ \rangle$: the possible two-particle states are
                $\left\{
                \displaystyle{\frac{a a^* + b b^*}{\sqrt{2}},\
                \frac{c c^* + d d^*}{\sqrt{2}},\
                \frac{e e^* + f f^*}{\sqrt{2}}} \right\}$
  and the corresponding scattering matrix reads
		\begin{equation}
		\label{M33_3}
			\begin{pmatrix}
				3a_1 & 2 b_3+c_3 & 2b_2+c_2 \\
				2b_3+c_3 & 3a_2 & 2b_1+c_1 \\
				2b_2+c_2 & 2b_1+c_1 & 3a_3 \\
			\end{pmatrix}.
		\end{equation}
\item $| 0,0,0,+,- \rangle$: the possible two-particle states are 
		$\left\{ 
		\left( ae^*+bf^* \right) \left/ \sqrt{2} \right.,\,
		\left( ea^*+fb^* \right) \left/ \sqrt{2} \right.
		\right\}$ and the corresponding scattering matrix reads
		\begin{equation}
			\begin{pmatrix}
				b_2+2c_2 & 3 d_2 \\
				3 d_2 & b_2+2c_2 \\
			\end{pmatrix}.
		\end{equation}
\item $| 0,0,0,-,+ \rangle$: the possible two-particle states are 
		$\left\{ 
		\left( ac^*+bd^* \right) \left/ \sqrt{2} \right.,\,
		\left( ca^*+db^* \right) \left/ \sqrt{2} \right.
		\right\}$ and the corresponding scattering matrix reads
		\begin{equation}
			\begin{pmatrix}
				b_3+2c_3 & 3 d_3 \\
				3 d_3 & b_3+2c_3 \\
			\end{pmatrix}.
		\end{equation}
\item $| 0,0,0,-,- \rangle$: the possible two-particle states are 
		$\left\{ 
		\left( ce^*+df^* \right) \left/ \sqrt{2} \right.,\,
		\left( ec^*+fd^* \right) \left/ \sqrt{2} \right.
		\right\}$ and the corresponding scattering matrix reads
		\begin{equation}
		\label{lastSM}
			\begin{pmatrix}
				b_1+2c_1 & 3 d_1e^{i\varepsilon} \\
				3 d_1e^{-i\varepsilon} & b_1+2c_1 \\
			\end{pmatrix}.
		\end{equation}
\end{itemize}

The eigenvalues of the $3\times 3$ matrices~\eqref{M33_1},
\eqref{M33_2}, and \eqref{M33_3} must be determined
numerically and then it must be required
that their moduli be smaller than $8 \pi$.
For the remaining matrices,
the perturbative unitarity constraints can be written as
\begin{equation}
\label{constssss}
\left| b_k \right| + \left| c_k \right| \leq  8\pi,\quad	
\left| b_k \right| + \left| d_k \right| \leq  8\pi,\quad	
\left| b_k+2c_k \right| + 3 \left| d_k \right| \leq  8\pi,
\quad \forall k=\{1,2,3\}.
\end{equation}
We have confirmed that both the scattering 
matrices~\eqref{M33_1}--\eqref{lastSM}
and inequalities~\eqref{constssss} are 
in agreement with the results derived in~\cite{Bento}.

\end{appendix}

\end{document}